\documentclass[a4paper,11pt]{article}
\pdfoutput=1 
\usepackage{jcappub}
\usepackage[english]{babel}
\usepackage{graphicx}
\usepackage{amsmath}
\usepackage{amssymb}
\usepackage{amsfonts}
\usepackage{lscape}
\usepackage{hyperref}
\usepackage{subfigure} 
\usepackage{xcolor}
\usepackage{bm}
\usepackage{tabularx}
\usepackage{mathtools}
\newcommand{\be}{\begin{equation}}
\newcommand{\ee}{\end{equation}}
\newcommand{\ba}{\begin{eqnarray}}
\newcommand{\ea}{\end{eqnarray}}

\renewcommand{\[}{\begin{equation}}
\renewcommand{\]}{\end{equation}}
\def\be{\begin{equation}}
\def\ee{\end{equation}}
\def\bea{\begin{eqnarray}}
\def\eea{\end{eqnarray}}
\def\eqi{\begin{equation}}
\def\eqf{\end{equation}}
\def\eqia{\begin{eqnarray}}
\def\eqfa{\end{eqnarray}}
\def\lcdm{$\Lambda$CDM }
\def\cs{\hat{c}_{\mathrm{s}}^{2}}
\def\wde{w_\mathrm{de}}

\def\dm{\delta_\mathrm{m}}
\def\Vm{V_\mathrm{m}}
\def\Omeff{\Omega_{\mathrm{m},0}^{\mathrm{eff}}}
\def\rhode{\rho_{\mathrm{de}}}
\def\omegab{\omega_{\rm{b}}}
\def\omegacdm{\omega_{\rm{cdm}}}
\def\ns{n_{\rm{s}}}
\def\As{\ln 10^{10}A_{\rm{s}}}

\definecolor{darkgreen}{rgb}{0,0.6,0}

\makeatother

\title{Holographic energy density, dark energy sound speed, and tensions in cosmological parameters: $H_0$ and $S_8$}

\author[a,1]{Wilmar Cardona,\note{Corresponding author.}}
\author[b]{M. A. Sabogal,}

\affiliation[a]{ICTP South American Institute for Fundamental Research \& Instituto de F\'isica Te\'orica, Universidade Estadual Paulista, 01140-070, S\~ao Paulo, Brazil}

\affiliation[b]{Programa de F\'isica, Universidad del Atl\'antico, Carrera 30 N\'umero 8-49, Puerto Colombia-Atl\'antico, Colombia}

\emailAdd{wilmar.cardona@unesp.br}
\emailAdd{msabogal@est.uniatlantico.edu.co}

\date{\today}

\abstract{
Interesting discrepancies in cosmological parameters are challenging the success of the $\Lambda$CDM model. Direct measurements of the Hubble constant $H_0$ using Cepheid variables and supernovae turn out to be higher than inferred from the Cosmic Microwave Background (CMB). Weak galaxy lensing surveys consistently report values of the strength of matter clustering $\sigma_8$ lower than values derived from the CMB in the context of $\Lambda$CDM. In this paper we address these discrepancies in cosmological parameters by considering Dark Energy (DE) as a fluid with evolving equation of state $\wde(z)$, constant sound speed squared $\cs$, and vanishing anisotropic stress $\sigma$. Our $\wde(z)$ is derived from the Holographic Principle and can consecutively exhibit radiation-like, matter-like, and DE-like behaviour, thus affecting the sound horizon and the comoving angular diameter distance, hence $H_0$. Here we show DE sound speed plays a part in the matter clustering behaviour through its effect on the evolution of the gravitational potential. We compute cosmological constraints using several data set combinations including primary CMB, CMB lensing, redshift-space-distortions, local distance-ladder, supernovae, and baryon acoustic oscillations. In our analysis we marginalise over $\cs$ and find $\cs=1$ is excluded at $\gtrsim 3\sigma$. For our baseline result including the whole data set we found $H_0$ and $\sigma_8$ in good agreement (within $\approx 2\sigma$) with low redshift probes. Our constraint for the baryon energy density $\omegab$ is however in $\approx 3\sigma$ tension with BBN constraints. We conclude evolving DE also having non-standard clustering properties [e.g., $\cs(z,k)$] might be relevant for the solution of current discrepancies in cosmological parameters.}
\makeatletter
\gdef\@fpheader{}
\makeatother
\DeclareUnicodeCharacter{2212}{-}
\begin{document}

\maketitle

\section{Introduction}
\label{Section:Introduction}

While the concordance model provides a reasonable, good phenomenological description of most astrophysical measurements \cite{Planck:2018vyg,DES:2018ekb,DES:2020mlx, Mossa:2020gjc}, it also becomes clear that our ignorance about the nature of Dark Matter as well as the so-called cosmological constant problem represent major drawbacks in the model. In addition, over the past years we have seen the emergence of pretty interesting discrepancies (e.g., the Hubble constant $H_0$, the strenght of matter clustering $\sigma_8$) in cosmological parameters whose understanding could reveal new physics disregarded in the standard cosmological model \cite{Abdalla:2022yfr,Riess:2019cxk,Freedman:2021ahq,Riess:2021jrx,Joudaki:2016mvz,Heymans:2020gsg,Philcox:2021kcw,Blanchard:2021dwr,DES:2021wwk,PhysRevD.91.103508,PhysRevLett.111.161301,Vagnozzi:2021gjh,Schoneberg:2021qvd,Gomez-Valent:2021cbe,Anchordoqui:2021gji,Rashkovetskyi:2021rwg,Bansal:2021dfh,Heisenberg:2022gqk,Dainotti:2021pqg,Dainotti:2022bzg,Murgia:2016ccp}.

Although Bayesian analyses show that the standard cosmological model \lcdm  performs better than its simplest alternatives \cite{Heavens:2017hkr}, there exists the possibility that more elaborate models could explain the shortcomings \lcdm is facing. Dynamical Dark Energy and Modified Gravity (MG) have become the two leading approaches when trying to explain the late-time accelerating universe \cite{Copeland:2006wr,Clifton:2011jh,Li:2011sd,Bamba:2012cp}. There is however no conclusive evidence for new Dark Energy (DE) fields or deviations from General Relativity \cite{PhysRevLett.116.221101,Collett:2018gpf,Planck:2015bue}. 

Within the wide spectrum of proposals to address the DE problem, there is a hypothesis known as the Holographic Principle (HP). Roughly speaking, the HP asserts that everything inside a region of space can be described by bits of information confined to the boundary \cite{tHooft:1993dmi,Susskind:1994vu,Hawking:1975vcx,Bekenstein:1973ur,Bousso:2002ju,Bousso:1999xy}. This non-extensive scaling would suggest that quantum field theory ceases to be valid in a large volume. Nevertheless, it is also true that the performance of local quantum field theory at describing particle phenomenology is quite remarkable. It turns out that a relationship between ultraviolet (UV) and infrared (IR) cut-offs of an effective quantum field theory could make these regimes compatible with each other \cite{Cohen:1998zx}. If $\rho$ is the quantum zero-point energy density associated to a UV cut-off,
the total energy in a region of size $L$ should not exceed the mass of a black hole of the same size, namely,
\be
L^3 \rho \leq L M_{\mathrm{p}}^2,
\label{eq:HP1}
\ee
where $M_{\mathrm{p}}$ is the reduced Planck mass. The largest, allowed IR cut-off $L_{\mathrm{IR}}$ saturates  the inequality \eqref{eq:HP1} so that the maximum energy density in the effective theory is given by
\be 
\rho = 3 \gamma^2 M_{\mathrm{p}}^2 L_{\mathrm{IR}}^{-2},
\label{eq:HP2}
\ee 
where $\gamma$ is an arbitrary parameter. The UV/IR relationship \eqref{eq:HP2} is a consequence of recognising that quantum field theory overestimates states. Moreover, it offers a possible way of understanding the cosmological constant problem \cite{Weinberg:1988cp,Sahni:1999gb,Carroll:2000fy}, one of the main shortcomings of the standard cosmological model $\Lambda$CDM.

Interestingly, the UV/IR relation \eqref{eq:HP2} has been widely applied in cosmology as an alternative to the cosmological constant causing the late-time accelerating expansion in the concordance model. These kinds of cosmological models are now known as Holographic Dark Energy (HDE) models (see \cite{Wang:2016och} for a review). In this context, the IR cut-off $L_{\mathrm{IR}}$ has a cosmological origin and various choices are found in the literature \cite{Nojiri:2005pu,Nojiri:2017opc,Granda:2008dk,Gao:2007ep,Nojiri:2021iko,Nojiri:2021jxf,Nojiri:2022dkr}. Despite being appealing as an alternative to $\Lambda$CDM, the HDE models investigated here are not derived from a Lagrangian which is a  disadvantage when studying the evolution of cosmological perturbations: since HDE models do not have a Lagrangian, we cannot derive equations of motion for linear order perturbations.\footnote{However, see Ref.~\cite{Lin_2021} for a relation between HDE and massive gravity theory that could provide a framework for investigating perturbations.} Nevertheless, fairly general theories relying on scalar and vector fields (e.g., scalar-vector-tensor theories \cite{Heisenberg:2018acv}) could provide background phenomenology matching HDE models while allowing the investigation of cosmological perturbations. Here we will adopt a phenomenological approach and assume the existence of a DE fluid having an evolving equation of state $\wde(a)$ derived from the UV/IR relation \eqref{eq:HP2}. As for the description of DE perturbations, we opt for a constant sound speed in the fluid rest-frame $\cs$ and vanishing anisotropic stress $\pi=0$.

In this work we want to determine whether or not HDE is viable given current astrophysical measurements. Although cosmological constraints have been computed for HDE models (see, for instance, \cite{Zhang:2009un,Xu:2010gg,Wang:2011km,Akhlaghi:2018knk,Hossienkhani:2021emv, Najafi:2022wjs,Cid:2020kpp,Malekjani:2018qcz,Cardenas:2013moa,PhysRevD.87.043525,Huang:2012gd,Zhang:2012qra,Fu:2011ab,PhysRevD.81.083523,Li:2009bn}), a few details have been overlooked. Firstly, while HDE models usually feature an evolving $\wde(a)$ which might cross the phantom divide $\wde=-1$, this behaviour is not properly addressed in the literature when also considering the evolution of perturbations. Here we will take it into consideration by using the Parameterized Post-Friedmann (PPF) formalism \cite{Fang:2008sn}. Secondly, when modelling DE perturbations, studies exist which a priori set $\cs$ to a constant value. However, this choice could bias cosmological constraints as it directly affects the clustering properties of DE. In our investigation we marginalise over $\cs$ and inquire about its phenomenological signatures in the context of HDE. Thirdly, with regard to cosmological constraints of HDE models, most studies focus on the background evolution and use only low red-shift data to constrain the parameter space fully disregarding the impact on earlier stages of the Universe. Here we fill this gap in the literature by also studying the impact of HDE on linear order perturbations. 

The manuscript is organised as follows. In Section \ref{Section:model} we set our notation, discuss the particular HDE model and explain its background phenomenology as well as the behaviour of linear order perturbations. In Sections \ref{Section:data-methodology}-\ref{section:results} we present and discuss results for cosmological constraints. Finally, in Section \ref{Section:conclusions} we give our conclusions.
    
\section{Theoretical framework and holographic dark energy model}
\label{Section:model}

The Einstein-Hilbert action reads 
\begin{equation}
S=\int d^{4} x \sqrt{-g}\left(\frac{R}{2 \kappa}+\mathcal{L}_{\mathrm{m}}\right),
\label{eq:EH-action}
\end{equation}
where   $\mathcal{L}_{\mathrm{m}}$ denotes the Lagrangian for any matter fields appearing in the theory, $g$ is the determinant of the metric $g_{\mu\nu}$, $R$ is the Ricci scalar and $\kappa \equiv 8 \pi G$ is a constant with $G$ being the bare Newton’s constant. By applying the Principle of Least Action we can derive the well known Einstein field equations 
\begin{equation}
R_{\mu \nu}-\frac{1}{2} R g_{\mu \nu} = \kappa T_{\mu\nu},
\label{eq:Einstein-Equations}
\end{equation}
where $R_ {\mu \nu}$ is the Ricci tensor and $T_{\mu\nu}$ is the energy-momentum tensor of matter fields.\footnote{Unless stated otherwise, throughout this paper we adopt the following conventions: speed of light $c=1$, $\tau$ is the conformal time, $\vec{x}$ denotes conformal comoving coordinates, and the metric signature is $(-+++)$. For a generic function $f$, $\dfrac{df}{d\tau}\equiv \dot{f}$ and $\dfrac{df}{da}\equiv f'$. Cosmic time $t$ and conformal time $\tau$ are related via $d\tau = dt/a(\tau)$.} Since observations and simulations indicate that on large enough scales the Universe is statistically homogeneous and isotropic also having vanishing curvature  \cite{Hogg:2004vw,Marinoni:2012ba,Ade:2015hxq,Planck:2018vyg}, here we will assume a flat, linearly perturbed  Friedmann-Lema\^{i}tre-Robertson-Walker metric (FLRW). In the conformal Newtonian gauge \cite{Ma-Bertschinger:1995asth}  
\begin{equation}
d s^{2}=a(\tau)^{2}\left[-(1+2 \psi(\vec{x}, \tau)) d \tau^{2}+(1-2 \phi(\vec{x}, \tau)) d \vec{x}^{2}\right], 
\label{eq:FLRW-metric}
\end{equation}
where $a(\tau)$ is the scale factor, and $\psi$, $\phi$ denote the gravitational potentials. As usual we will consider the material content as described by a perfect fluid with energy-momentum tensor
\begin{equation}
T_{\nu}^{\mu}= P_{\mathrm{fld}} \delta_{\nu}^{\mu}+(\rho_{\mathrm{fld}} + P_{\mathrm{fld}}) U^{\mu},
\label{eq:energy-momentum-tensor}
\end{equation}
where $\rho_{\mathrm{fld}}$, $P_{\mathrm{fld}}$, and $U^{\mu}$ respectively denote the energy density, pressure, and four-velocity vector of the fluid. At first order the four-velocity vector is given by $U^{\mu} = a(\tau)^{-1}\left( 1-\psi,\vec{u}\right)$, which satisfies $U^{\mu}U_{\mu}=-1$, with $\vec{u}=\dot{\vec{x}}$. Taking into account linear perturbations, the elements of the energy-momentum tensor are given by
\begin{align}
T_{0}^{0} &=-(\bar{\rho}_{\mathrm{fld}}+\delta \rho_{\mathrm{fld}}), \label{eq:energy-momentum-tensor-perturbed-1} \\ 
T_{i}^{0} &=(\bar{\rho}_{\mathrm{fld}}+\bar{P}_{\mathrm{fld}}) u_{i},\\
T_{j}^{i} &=(\bar{P}_{\mathrm{fld}}+\delta P_{\mathrm{fld}}) \delta_{j}^{i}+\Sigma_{j}^{i},
\label{eq:energy-momentum-tensor-perturbed}    
\end{align}
where $\bar{\rho}_{\mathrm{fld}}$, $\bar{P}_{\mathrm{fld}}$ are background quantities and only depend on time. The perturbations $\delta \rho_{\mathrm{fld}}$, $\delta P_{\mathrm{fld}}$, $\Sigma_{j}^{i}$ depend on $(\vec{x},\tau)$. The anisotropic stress tensor of the fluid is defined as $\Sigma_{j}^{i} \equiv T_{j}^{i} - \delta_{j}^{i} T_{k}^{k}/3$.

\subsection{Background}

From the time-time component of Eq. \eqref{eq:Einstein-Equations} and using the unperturbed (i.e., $\psi=\phi=0$) FLRW metric \eqref{eq:FLRW-metric}, we obtain
\begin{equation}
H^{2}=\frac{\kappa}{3}(\rho_{\mathrm{r}} + \rho_{\mathrm{m}} + \rho_{\mathrm{de}}),
\label{eq:Friedmann1General}
\end{equation}
where the Hubble parameter $H \equiv \dfrac{1}{a(t)} \dfrac{da}{dt} $, and $\rho_{\mathrm{de}}$, $\rho_{\mathrm{r}}$,  $\rho_{\mathrm{m}}$   respectively denote DE, radiation, and matter energy densities. While radiation and matter will be taken into account as in the standard cosmological model $\Lambda$CDM, we will consider DE as a fluid with energy density given by  \eqref{eq:HP2}. We choose the so-called GO cut-off \cite{Granda:2008dk} 
\begin{equation}
    L_{\mathrm{IR}}^{-2} \equiv \alpha H^{2}+\beta \dfrac{dH}{dt}
\label{eq:GO-cutoff}    
\end{equation}
where $\alpha$ and $\beta$ are dimensionless constants. Eqs. \eqref{eq:HP2} and \eqref{eq:GO-cutoff} allow us to define a HDE density
\begin{equation}
\rho_{\mathrm{de}}= \frac{3}{\kappa}  \left(\alpha H^{2}+\beta \dfrac{dH}{dt}\right),
\label{eq:GO-HDE}
\end{equation}
where the constant $\gamma$ was absorbed by $\alpha$ and $\beta$. Taking into account Eq. \eqref{eq:GO-HDE}, we can rewrite the Friedmann equation \eqref{eq:Friedmann1General} as\footnote{As it is usual, we define the density parameters $\Omega_{i,0} \equiv \frac{\kappa  }{3 H_{0}^{2}}\rho_{i,0}$ and use $ \dfrac{d}{dt} = a H \dfrac{d}{da}$.}
\begin{equation}
H^{2} = \Omega_{\mathrm{r},0} H_{0}^{2} a^{-4} + \Omega_{\mathrm{m},0} H_{0}^{2} a^{-3}  + \left(\alpha H^{2}+ \beta \frac{a}{2} \frac{d H^{2}}{da}\right). 
\label{eq:differential-HDE}
\end{equation}
We define $E^{2} \equiv \frac{H^{2}}{H_{0}^{2}}$ and find an analytical solution for the differential equation \eqref{eq:differential-HDE}  given by
\be 
E^{2}(a) = \Omega_{\mathrm{r},0}^{\mathrm{eff}} a^{-4} + \Omega_{\mathrm{m},0}^{\mathrm{eff}} a^{-3}  + \Omega_{\mathrm{de},0}^{\mathrm{eff}} a^{ \frac{-2(\alpha-1)}{\beta}},
\label{eq:HDE-solution-H}
\ee
where
\be 
\Omega_{\mathrm{r},0}^{\mathrm{eff}} \equiv \left( 1 + \frac{(\alpha-2 \beta)}{(1-\alpha+2 \beta)} \right) \Omega_{\mathrm{r},0},
\label{eq:omega-radiation-effective-today}
\ee 
\be
\Omega_{\mathrm{m},0}^{\mathrm{eff}} \equiv \left( 1 + \frac{(2 \alpha-3 \beta)}{(2-2 \alpha +3 \beta)}  \right) \Omega_{\mathrm{m},0},
\label{eq:omega-matter-effective-today}
\ee
\be 
\Omega_{\mathrm{de},0}^{\mathrm{eff}} \equiv \left(1-\frac{2 \Omega_{\mathrm{m},0}}{(2-2 \alpha +3 \beta)}  -\frac{ \Omega_{\mathrm{r},0} }{(1-\alpha+2 \beta)} \right),
\label{eq:omega-de-today}
\ee
and the effective parameter densities satisfy  $\Omega_{\mathrm{m},0}^{\mathrm{eff}} +  \Omega_{\mathrm{r},0}^{\mathrm{eff}} +  \Omega_{\mathrm{de},0}^{\mathrm{eff}} = 1$. Note that the HDE parameter density reads
\be 
\Omega_{\rm{de}} = \left(\frac{\alpha-2 \beta}{1-\alpha+2 \beta} \right) \Omega_{\mathrm{r},0} a^{-4} + \left(\frac{2 \alpha-3 \beta}{2-2 \alpha +3 \beta}  \right) \Omega_{\mathrm{m},0} a^{-3} + \Omega_{\mathrm{de},0}^{\mathrm{eff}} a^{ \frac{-2(\alpha-1)}{\beta}}.
\label{eq:Omega-de}
\ee
Assuming a barotropic fluid with $P_\mathrm{de} = w_\mathrm{de} \rho_\mathrm{de}$, from the condition for energy conservation 
\begin{equation}
    \dfrac{d\rho_{\mathrm{de}}}{dt} + 3 H \rho_{\mathrm{de}} \left( 1 + w_{\mathrm{de}} \right)=0
    \label{eq:energy-conservation}
\end{equation}
and Eqs. \eqref{eq:GO-HDE} and \eqref{eq:HDE-solution-H}, we can derive the equation of state for our DE fluid
\be 
w_{\mathrm{de}}(a)= \frac{\left(\frac{2 \alpha-3 \beta-2}{3 \beta}\right)    \Omega_{\mathrm{de},0}^{\mathrm{eff}}
a^{ \frac{-2(\alpha-1)}{\beta}} + \left(\frac{2 \beta-\alpha}{3\alpha-6 \beta-3}\right)   \Omega_{\mathrm{r},0} a^{-4} }{\left(\frac{2 \alpha-3 \beta}{2-2 \alpha +3 \beta}\right)  \Omega_{\mathrm{m},0} a^{-3} +\left(\frac{\alpha-2 \beta}{1-\alpha+2 \beta}\right) \Omega_{\mathrm{r},0} a^{-4} + \Omega_{\mathrm{de},0}^{\mathrm{eff}} a^{ \frac{-2(\alpha-1)}{\beta}} }. 
\label{eq:EOS_GO}
\ee

\begin{figure}[http]
\begin{center}
\includegraphics[scale=0.74]{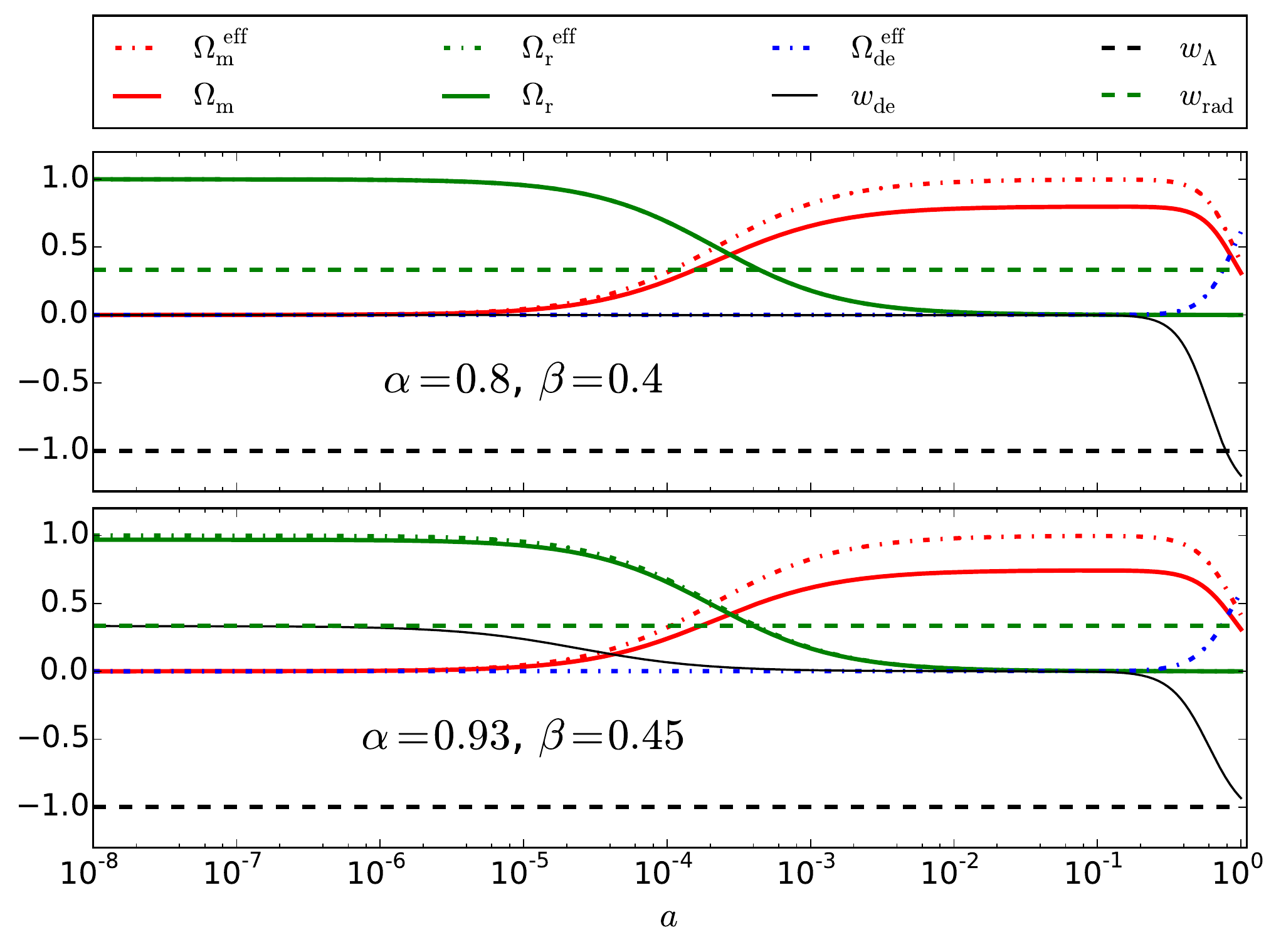}
\caption{Evolution of parameter densities and HDE equation of state $\wde$ for the HDE model. Note that $\wde$ can cross the phantom divide.  Here we use $\Omega_{m,0} = 0$.$31$ and $\Omega_{r,0} = 8$.$5 \times 10^{-5}$. While in the lower panel ($\alpha>2\beta$) the HDE equation of state has a radiation-like behaviour at early times, in the upper panel ($\alpha=2\beta$) $\wde$ is matter-like when the universe is under radiation dominance.}
\label{fig:phenomenology-background-1}
\end{center}
\end{figure}

Figure \ref{fig:phenomenology-background-1} shows the evolution of parameter densities as well as the HDE equation of state $w_{\mathrm{de}}(a)$ in Eq. \eqref{eq:EOS_GO}. It becomes clear that when $\alpha > 2\beta$ the HDE equation of state evolves from radiation-like [$w_{\mathrm{de}}(a) \approx 1/3$] to pressure-less matter-like [$w_{\mathrm{de}}(a) \approx 0$] until reaching a DE-like [$w_{\mathrm{de}}(a) < -1/3$] behaviour at late times. Consequently, a non-vanishing HDE \eqref{eq:GO-HDE} can effectively add both pressure-less matter and radiation to the cosmological model [see Eqs. \eqref{eq:omega-radiation-effective-today}-\eqref{eq:omega-matter-effective-today}]. While for the case where $\alpha=2\beta$ there is no radiation-like behaviour of HDE in the early universe, HDE contributes to the effective matter parameter density in the matter dominated epoch.
Since in this work we focus on a possible explanation for the late time accelerating expansion of the Universe and its relation with the formation of structures, we constrain the HDE model to satisfy $\Omega_{\mathrm{r},0}^{\mathrm{eff}}=\Omega_{\mathrm{r},0}$ later when computing cosmological constraints. From Eq. \eqref{eq:omega-radiation-effective-today}, the latter is fulfilled for $\alpha=2\beta$. In this way we make sure that the early universe is described as in the $\Lambda$CDM model. Later in Section \ref{section:results} we will expand on this constraint in relation with previous works.

With the constraint $\alpha=2\beta$, the HDE density in Eq. \eqref{eq:GO-HDE} becomes the well known Ricci Dark Energy (RDE) which only has a single free parameter. Then, the normalised Hubble parameter \eqref{eq:HDE-solution-H} and the HDE equation of state \eqref{eq:EOS_GO} are simplified
\be
E^{2}(a) = \Omega_{\mathrm{r},0} a^{-4} + \left(1 + \dfrac{\alpha}{4-\alpha} \right)  \Omega_{\mathrm{m},0} a^{-3} +  \left(1- \dfrac{\Omega_{\mathrm{m},0}}{1-\dfrac{\alpha}{4}} -  \Omega_{\mathrm{r},0} \right) a^{ \frac{4-4\alpha}{\alpha}}, 
\label{eq:HDE-solution-H-Ricci}
\ee
\be 
w_{\mathrm{de}}(a)=  \frac{\alpha - 4 }{3 \alpha \left(1-\frac{\alpha \Omega_{\mathrm{m},0} }{(\alpha - 4) (1-\Omega_{\mathrm{r},0} ) + 4 \Omega_{\mathrm{m},0}}   a^{ \frac{\alpha-4}{\alpha}}  \right) }. 
\label{eq:Ricci-EoS}
\ee
From Eq. \eqref{eq:HDE-solution-H-Ricci} we can easily extract the HDE density
\be
\tilde{\rho}_{\mathrm{de}}(a) = \frac{ \alpha }{(4- \alpha)}  \Omega_{\mathrm{m},0} a^{-3}  +\left(1-\frac{ 4 }{(4- \alpha)} \Omega_{\mathrm{m},0} - \Omega_{\mathrm{r},0} \right)a^{ \frac{4-4\alpha}{\alpha}}, 
\label{eq:Ricci-density}
\ee
where $\tilde{\rho}_{\mathrm{de}} \equiv \frac{\rho_{\mathrm{de}}}{H_0^2}$. Moreover, from the conservation equation \eqref{eq:energy-conservation} we can derive an expression for the HDE pressure $\tilde{P}_{\mathrm{de}} \equiv \frac{P_{\mathrm{de}}}{H_0^2}$  
\be
\tilde{P}_{\mathrm{de}}(a)= - \frac{4-\alpha}{3 \alpha} \left(1-\frac{4}{(4-\alpha)} \Omega_{\mathrm{m},0} \right) a^{ \frac{4-4\alpha}{\alpha}}, 
\ee
which in turn allows us to compute  the adiabatic sound speed squared for the DE fluid \begin{align}
 c_{a}^{2} & \equiv  \dfrac{dP_\mathrm{de}}{d\rho_\mathrm{de}} = \wde - \dfrac{\dot{\wde} }{3 H (1+\wde)} = \wde - \dfrac{ \wde^{\prime} a }{3 (1+\wde)} \nonumber \\ 
 & = \frac{ 4(\alpha - 4) }{3 \alpha \left(4-\frac{ 3 \alpha^{2} \Omega_{\mathrm{m},0} }{ (\alpha-1)\left( (\alpha - 4) (1-\Omega_{\mathrm{r},0} ) + 4 \Omega_{\mathrm{m},0} \right) }   a^{ \frac{\alpha-4}{\alpha}}  \right) }. 
\label{eq:DE-adiabatic-sound-speed-squared}
\end{align}
We will use the previous expressions to derive simplified, approximate solutions for the DE  perturbations during matter dominance in the next section.

\subsection{First order perturbations}

In this work we are interested in computing statistical properties of observables such as the CMB angular power spectra and the matter power spectrum as predicted by the HDE model. Here we limit ourselves to first order scalar perturbations and therefore we need to solve the differential equations governing the linearised Einstein field equations \eqref{eq:Einstein-Equations}. Using the conformal Newtonian gauge \eqref{eq:FLRW-metric} and taking into account a general fluid \eqref{eq:energy-momentum-tensor-perturbed-1}-\eqref{eq:energy-momentum-tensor-perturbed}, we obtain
\begin{align}
k^{2} \phi+3 \frac{\dot{a}}{a}\left(\dot{\phi}+\frac{\dot{a}}{a} \psi\right)&=4 \pi G a^{2} \delta T_{0}^{0}, \label{eq:Ep1} \\
k^{2}\left(\dot{\phi}+\frac{\dot{a}}{a} \psi\right)&=4 \pi G a^{2}(\bar{\rho}_\mathrm{fld} + \bar{P}_\mathrm{fld}) \theta_\mathrm{fld} ,\label{eq:Ep2} \\
\ddot{\phi}+\frac{\dot{a}}{a}(\dot{\psi}+2 \dot{\phi})+\left(2 \frac{\ddot{a}}{a}-\frac{\dot{a}^{2}}{a^{2}}\right) \psi+\frac{k^{2}}{3}(\phi-\psi)&=\frac{4 \pi}{3} G a^{2} \delta T_{i}^{i}, \label{eq:Ep3} \\
k^{2}(\phi-\psi)&=12 \pi G a^{2}(\bar{\rho}_\mathrm{fld} +\bar{P}_\mathrm{fld}) \sigma_\mathrm{fld}, \label{eq:Ep4}
\end{align}
where $k$ is the wavenumber, the divergence of the velocity field is defined as $\theta_\mathrm{fld} \equiv i k^{j} u_{j}$, and $\sigma_\mathrm{fld}$ is the anisotropic stress. In the cases where the universe is regarded as composed by several fluids, the right-hand side in Eqs. \eqref{eq:Ep1}-\eqref{eq:Ep4} is intended to be a sum over all species (e.g., radiation, matter, dark energy).
From the conservation of energy-momentum $\left( \nabla_{\mu} T^{\mu \nu} = 0 \right)$ for a single fluid we obtain
\begin{align}
\dot{\delta}_\mathrm{fld}&=- V_\mathrm{fld} + 3(1+w_\mathrm{fld}) \dot{\phi} - 3 \frac{\dot{a}}{a}\left( \dfrac{\delta P_\mathrm{fld}}{\bar{\rho}_\mathrm{fld}}-w_\mathrm{fld} \delta_\mathrm{fld} \right), \label{eq:delta1}\\
\dot{V}_\mathrm{fld}&=-\frac{\dot{a}}{a}(1-3 w_\mathrm{fld})V_\mathrm{fld} + \dfrac{\delta P_\mathrm{fld}}{\bar{\rho}_\mathrm{fld}} k^{2}  + k^{2} (1+w_\mathrm{fld}) \phi,\label{eq:v2}
\end{align}
where we have used the scalar velocity perturbation $V_{\mathrm{fld}} \equiv i k_{j} T_{0}^{j}/\bar{\rho} =(1 + w_{\mathrm{fld}})\theta_{\mathrm{fld}}$ and disregarded anisotropic stress.

Since we are interested in the observational signatures of the HDE cosmological model at late-times, in the remainder of this section we will discuss the behaviour of matter and DE perturbations starting our analysis in the matter dominated epoch. We consider pressure-less matter with $w_\mathrm{m}=0$, $\delta P_\mathrm{m}=0$, $\sigma_\mathrm{m}=0$ and assume HDE as a fluid having DE equation of state \eqref{eq:EOS_GO}, pressure perturbation $\delta P_{\rm{de}}$, and vanishing anisotropic stress $\sigma_{\rm{de}} = 0$. The latter and Eq. \eqref{eq:Ep4} imply that at late-times the gravitational potentials $\phi=\psi$.

We parameterise the DE pressure perturbation as
\begin{equation}
\frac{\delta P_{\rm{de}}}{\bar{\rho}_\mathrm{de}} = \hat{c}_{\mathrm{s}}^{2} \delta +\frac{3 a H\left(\hat{c}_{\mathrm{s}}^{2}-c_{a}^{2}\right)}{k^{2}}  V
\label{eq:deltaP}
\end{equation}
where $\hat{c}_{\mathrm{s}}^{2}$ is the DE sound speed squared in the rest-frame, $\delta$ is the DE density perturbation, and $V$ is the DE velocity perturbation. Using \eqref{eq:deltaP}, we rewrite Eqs. \eqref{eq:delta1}-\eqref{eq:v2} for the DE perturbations
\begin{equation}
\delta^{\prime} = -\frac{V}{H a^{2}}\left(1+\frac{9 a^{2} H^{2}\left(\cs-\wde\right)}{k^{2}} +\frac{3 a^{3} H^{2} \wde^{\prime} }{k^{2} (1+\wde)}  \right) -\frac{3}{a}\left(\cs-\wde\right) \delta+3(1+\wde) \phi^{\prime},
\label{eq:DGO}
\end{equation}
\begin{equation}
V^{\prime} = -\left(1-3 \cs - \dfrac{a \wde^{\prime}}{ (1+\wde) } \right) \frac{V}{a}+\frac{k^{2} \cs}{H a^{2}} \delta+(1+\wde) \frac{k^{2}}{H a^{2}} \phi,
\label{eq:VGO}
\end{equation}
whereas for matter the perturbation equations \eqref{eq:delta1}-\eqref{eq:v2} become
\begin{equation}
\dm^{\prime} =-\frac{\Vm}{H a^{2}}+3 \phi^{\prime},
\label{eq:Dm}
\end{equation}
\begin{equation}
\Vm^{\prime} =-\frac{\Vm}{a}+\frac{k^{2}}{H a^{2}} \phi.
\label{eq:Vm}
\end{equation}

Note we can combine Eqs. \eqref{eq:Ep1}-\eqref{eq:Ep2} and obtain
\begin{equation}
k^{2} \phi=-4 \pi G a^{2} \sum_{j} \rho_{j}\left(\delta_{j}+\frac{3 a H}{k^{2}} V_{j}\right) \hspace{0.1cm}.
\label{eq:potential}
\end{equation}

\subsubsection{Matter dominance}

Here we will work out the solution for the system of differential equations \eqref{eq:DGO}-\eqref{eq:Vm} governing the evolution of matter and DE perturbations. We focus on late times starting from the epoch when matter becomes dominant so that we can safely neglect radiation in the model. Since we are interested in analytical, approximate solutions, to simplify our problem we assume that during Matter Dominance (MD) the Hubble parameter is 
\begin{equation}
H^{2}= H_{0}^{2} \Omeff a^{-3}.
\label{eq:H-MD}
\end{equation}
For the standard cosmological model \lcdm only matter contributes to the pressure perturbation in the right-hand side of Eq.~\eqref{eq:Ep3}, hence the solution for the gravitational potential $\phi$ takes on a constant value under MD. The situation is different for the HDE we investigate here because the DE fluid might have not negligible contributions to the pressure perturbation. The latter is parameterised by Eq.~\eqref{eq:deltaP} and therefore we identify two situations where $\phi$ is constant as in \lcdm: i) $\alpha=2\beta$ ($c_{\rm{a}}^2=0$) and $\cs=0$ so that $\delta P_{\rm{de}}=0$; ii) the more general scenario where $\alpha$, $\beta$ are independent parameters ($c_{\rm{a}}^2\approx0$) and $\cs=0$ implying $\delta P_{\rm{de}}\approx 0$.

We could only find analytical solutions for the perturbations when the gravitational potential takes on a constant value $\phi_0$. Therefore, using Eq.~\eqref{eq:H-MD} in Eqs.~\eqref{eq:DGO}-\eqref{eq:Vm}, we find the solutions for both matter and DE perturbations in MD  
\begin{equation}
\dm = \delta_{0}\left(a + 3\frac{H_{0}^{2} \Omeff }{k^{2}}\right), 
\label{eq:dm-MD-2}
\end{equation}
\begin{equation}
\Vm = -\delta_{0} H_{0} \sqrt{\Omeff} a^{1/ 2},
\label{eq:Vm-MD-2}
\end{equation}
\begin{equation}
\delta = \dm,
\label{eq:d-DE-MD}
\end{equation}
\begin{equation}
V = \Vm,
\label{eq:V-DE-MD}
\end{equation}
where
\begin{align}
\delta_0 \equiv &  -\frac{2 k^2 \phi_0}{3 H_0^2 \Omeff}.
\end{align}
Note that the potential $\phi$ in Eq.~\eqref{eq:potential} also receives contributions from the DE fluid in MD: DE perturbations behave as matter perturbations [see Eqs.~\eqref{eq:d-DE-MD}-\eqref{eq:V-DE-MD}] and the background DE density is not negligible [see Eq.~\eqref{eq:Omega-de}] under MD.

\begin{figure}[http]
\centering
\includegraphics[scale=0.85]{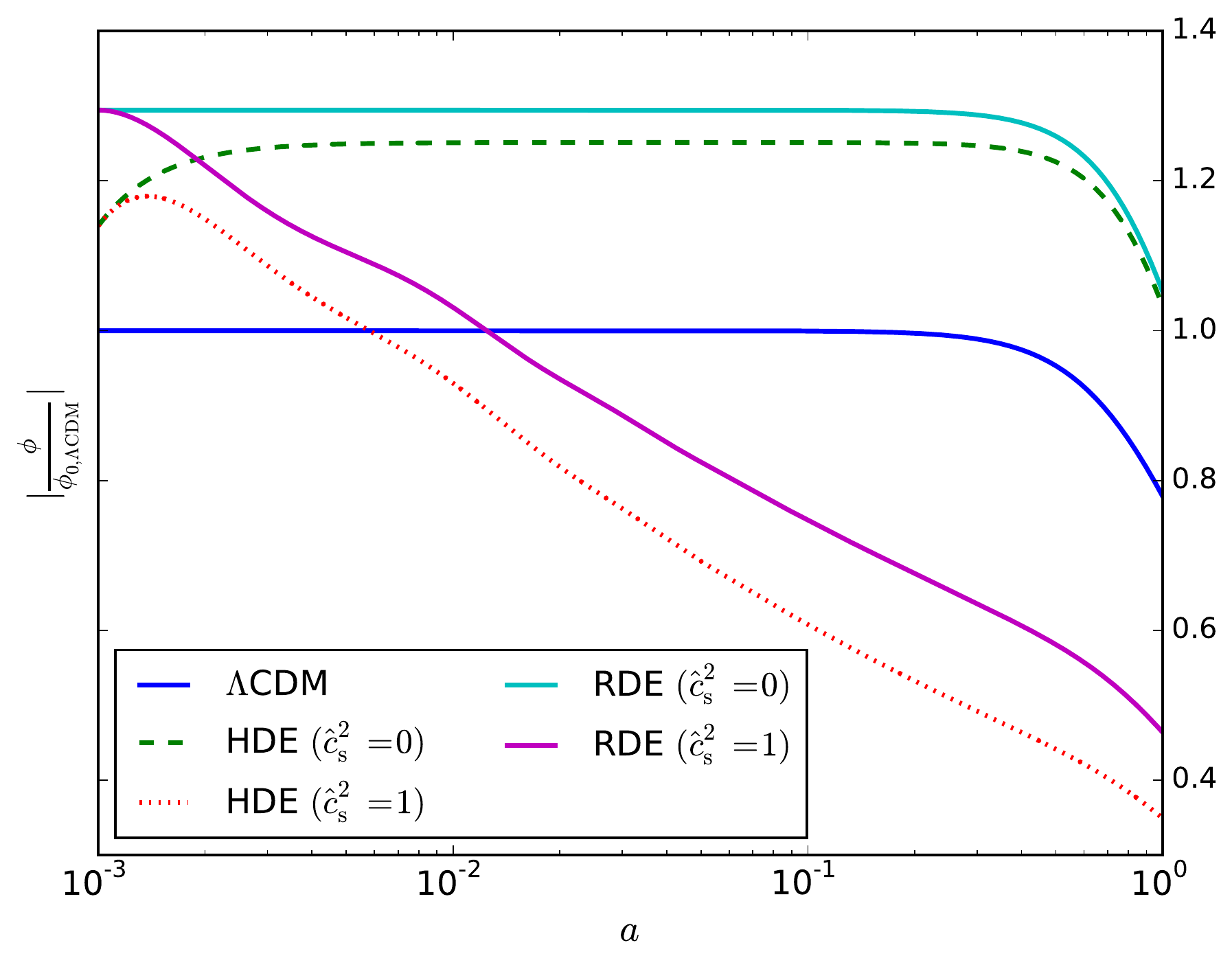}
\caption{Evolution of the gravitational potential in $\Lambda$CDM, HDE, and RDE cosmological models. All the plots are normalised by the initial value in $\Lambda$CDM. Common cosmological parameters used to numerically solve the system of differential equations are: $\Omega_{\rm{r},0}=0$, $\Omega_{\rm{m},0}=0.3$, $H_0=70\, \rm{km}\,\rm{s}^{-1}\,\rm{Mpc}^{-1}$, $k=25 H_0$, $\delta_0=1$; for RDE $\alpha=0.91$; for HDE $\alpha=0.88$ and $\beta=0.39$.}
\label{fig:gravitational-potential}
\end{figure}

We numerically solved the system of differential equations \eqref{eq:DGO}-\eqref{eq:Vm} also taking into account the expression \eqref{eq:potential} for the gravitational potential. The latter is depicted in Figure \ref{fig:gravitational-potential} for the standard cosmological model \lcdm along with solutions for HDE and RDE models. It becomes clear that while models having $\cs=1$ exhibit a varying gravitational potential during MD, models with a vanishing sound speed present a behaviour similar to $\Lambda$CDM, namely, a constant $\phi$. Note that for the given set of cosmological parameters, HDE ($\cs=0$) enters later than RDE ($\cs=0$) the regime of MD; the small variation of $\phi$ in this case is due to a non-vanishing adiabatic sound speed affecting the DE pressure perturbation, hence the gravitational potential.

\begin{figure}[http]
\centering
\includegraphics[scale=0.85]{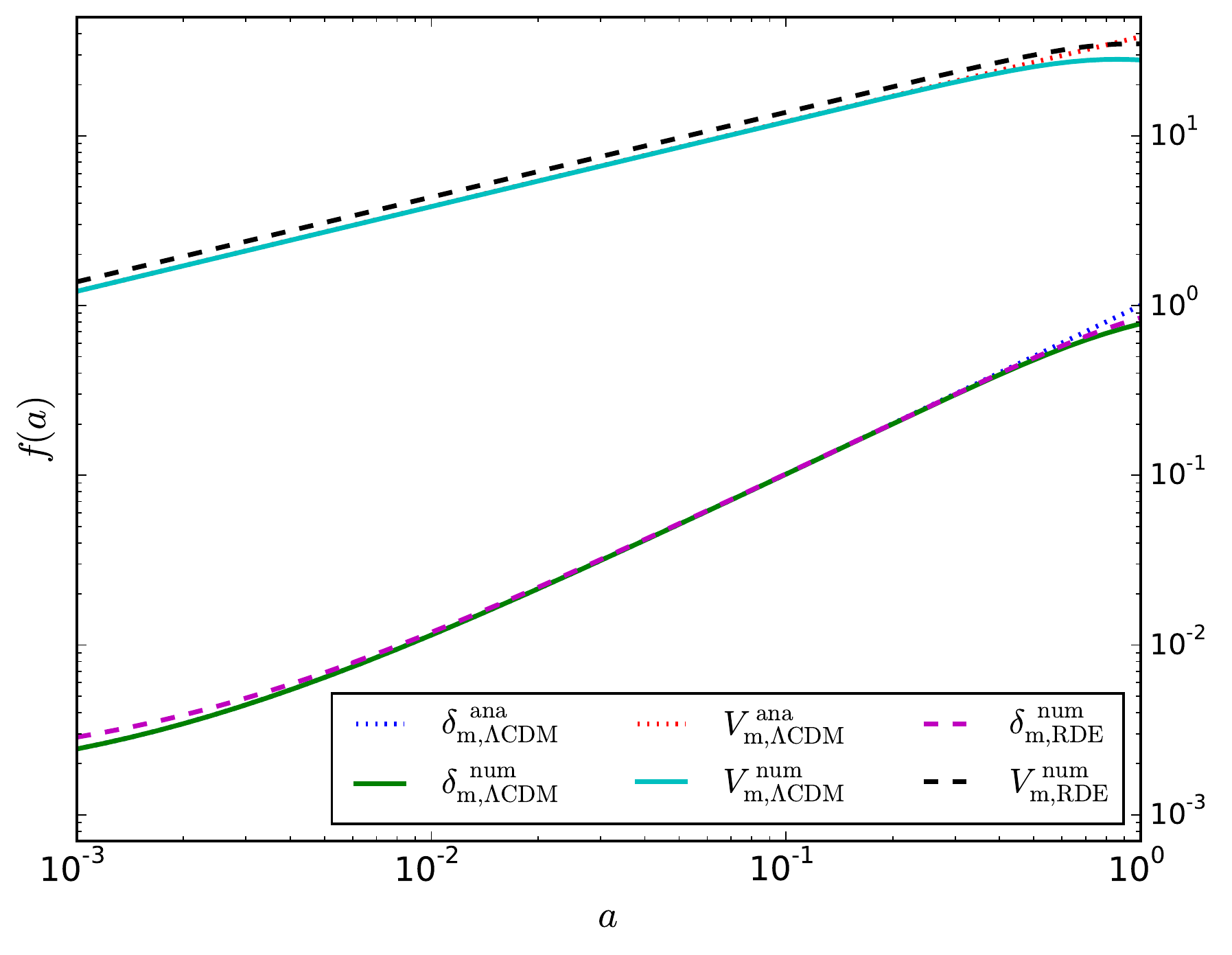}
\caption{Evolution of density and velocity matter perturbations in the \lcdm and RDE ($\cs=0$) model. We show the numerical solutions (solid and dashed curves) as well as the analytical solutions (dotted curves) in MD for the standard cosmological model. Cosmological parameters are as specified in Fig.~\ref{fig:gravitational-potential}.}
\label{fig:matter-perturbations}
\end{figure}

In Figure \ref{fig:matter-perturbations} we show the numerical solutions for the matter perturbations. We display solutions for the concordance model \lcdm and for the RDE ($\cs=0$) model. In the case of \lcdm we also show the well-known analytical solutions in the MD epoch. Differences in the solutions of density perturbations arise mainly before horizon crossing and when DE becomes dominant. This is indeed due to the modifications introduced by the RDE via $\Omeff$. With regard to the velocity perturbations, we can see that even though the dependence with the scale factor is the same in the two models, the RDE ($\cs=0$) solution fully  disagrees with the standard model due to $\Omeff$.

\begin{figure}
\centering
\includegraphics[scale=0.8]{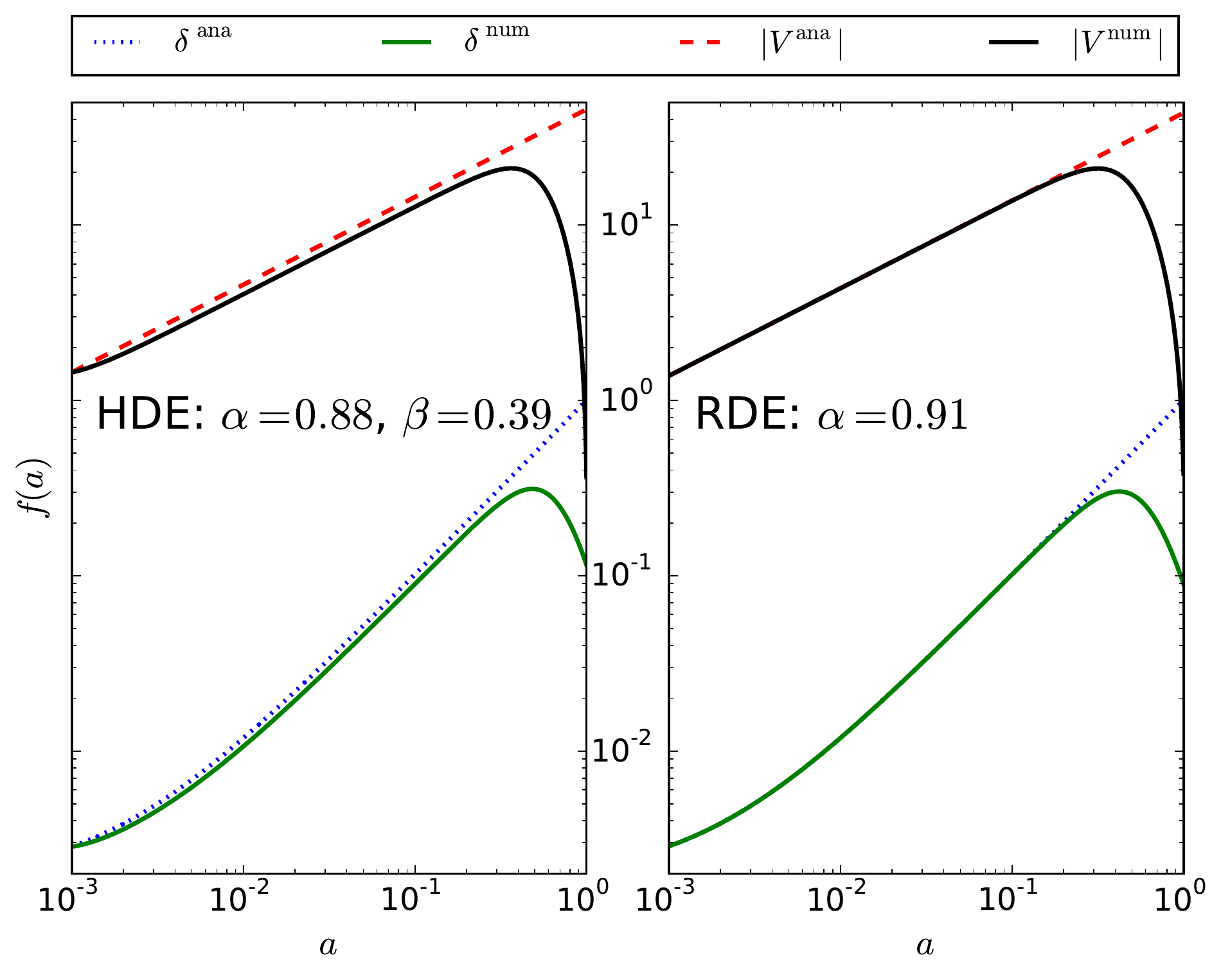}
\caption{The evolution of DE perturbations in HDE (left panel) and RDE (right panel). We compare analytical (dashed and dotted curves) and numerical (solid curves) solutions for the system of differential equations \eqref{eq:DGO}-\eqref{eq:Vm}. Cosmological parameters are as specified in Fig.~\ref{fig:gravitational-potential}.}
\label{fig:DE-perturbations}
\end{figure}

In Figure \ref{fig:DE-perturbations} we compare the numerical solutions for DE perturbations against the analytical, approximate solutions \eqref{eq:d-DE-MD}-\eqref{eq:V-DE-MD} valid in the MD regime. We carry out the comparison for HDE ($\cs=0$) in the left panel and RDE ($\cs=0$) in the right panel. While analytical solutions describe pretty well the behaviour of $\delta$ and $V$ in the RDE ($\cs=0$) model during MD, we find disagreement for the HDE ($\cs=0$) model. This is mainly due to the fact that in the latter full MD starts later than in RDE ($\cs=0$) for the given set of cosmological parameters.

\subsubsection{Implementation in Boltzmann solver}

Thus far in our treatment of perturbations, we regarded DE perturbations as governed by fluid equations. We used a prescription for the pressure perturbation given by \eqref{eq:deltaP}. However, this approach has problems (e.g., divergences, instabilities) when the DE equation of state crosses the phantom divide which turns out to be the case in the HDE model we investigate here (see Fig.~\ref{fig:phenomenology-background-1}). Not allowing cosmological models to cross $\wde=-1$ could introduce unwanted bias in our modelling since the phantom divide is currently a crucial point \cite{Kunz:2006PhyRev}. By requiring strict energy and momentum conservation, the PPF formalism (also implemented in the Boltzmann solver \texttt{CLASS}) can deal with “smooth” DE crossing the phantom divide \cite{Fang:2008sn}.      

The PPF description of DE replaces the density and momentum components with a single joint dynamical variable 
\begin{equation}
\Gamma\equiv -\frac{4 \pi G a^{2}}{k^{2} c_{K}} \delta \hat{\rho}_{\mathrm{de}}, 
\end{equation}
thus reducing closure conditions, but requiring strict conservation of energy and momentum in its equation of motion. Here $c_{K}=1-3K/k^{2}$, where $K$ is the space-time curvature that we set to $K=0$. The evolution equation for $\Gamma$ is given by
\begin{equation}
\left(1+c_{\Gamma}^{2} k_{H}^{2}\right)\left[\frac{\dot{\Gamma}}{H}+\Gamma+c_{\Gamma}^{2} k_{H}^{2} \Gamma\right]=S,
\end{equation}
where $c_{\Gamma} \equiv 0.4\,\hat{c}_{\mathrm{s}}$  calibrates the scale of the transition, $k_{H}=k^{2}/aH$ and   
\begin{equation}
S=\frac{ \dot{a} }{a} \frac{4 \pi G}{H^{2}} \rho_{\mathrm{de}} \left( 1+ \wde \right) \frac{\theta_{T} }{k^{2}} 
\end{equation}
where the subscript T denotes all species except dark energy. 

\begin{figure}[http]
\centering 
\includegraphics[scale=0.74]{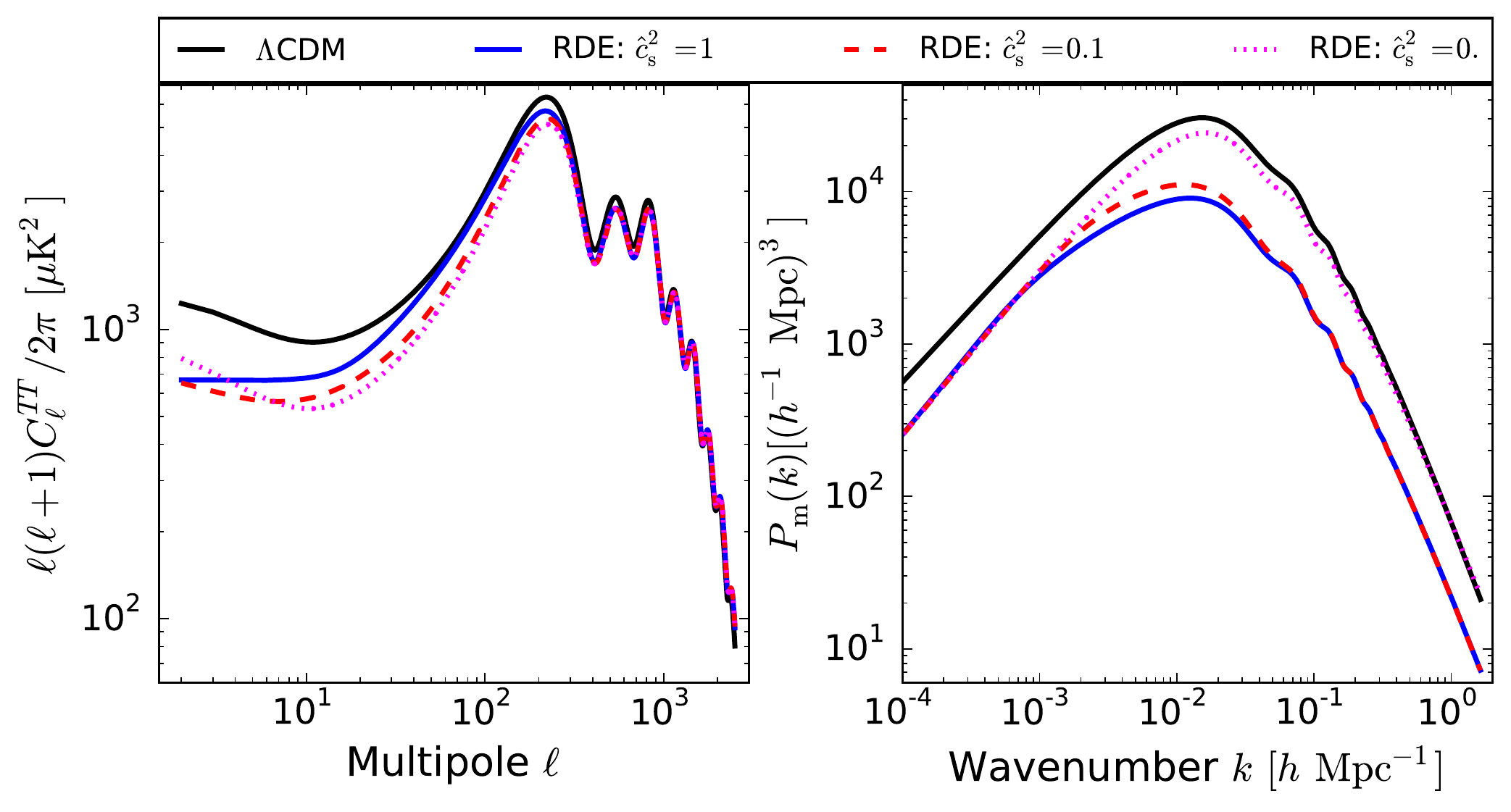}
\caption{CMB temperature angular power spectrum $C_{l}^{TT}$ and  matter power spectrum $P(k,z=0)$ for different values of $\cs$. We fix other cosmological parameters to the best fit values reported in Table 1 [\texttt{WMAP+BAO+SNIa}] of Ref.~\cite{Wang:2011km}, namely: for RDE $\omega_{\rm{b}} = 0.0241$, $\omega_{\rm{cdm}} = 0.1086$, $H_{0} = 72.26\, \rm{km}\,\rm{s}^{-1}\,\rm{Mpc}^{-1}$, $n_{\rm{s}}= 1.0871$, $\ln10^{10} A_{\rm{s}} = 3.122$, $\tau= 0.1382$ and $\alpha = 0.6904$ (note that authors in Ref.~\cite{Wang:2011km} actually report $\beta=0.3452$); for \lcdm $\omega_{\rm{b}} = 0.0226$, $\omega_{\rm{cdm}} = 0.1123$, $H_{0} = 70.38\, \rm{km}\,\rm{s}^{-1}\,\rm{Mpc}^{-1}$, $n_{\rm{s}}= 0.9691$, $\ln10^{10} A_{\rm{s}} = 3.180$, $\tau= 0.0877$. }
\label{fig_PRD_Observables_best}
\end{figure}

Figure \ref{fig_PRD_Observables_best} shows the output of our implementation in \texttt{CLASS} for a RDE model crossing the phantom divide. We use the best fit cosmological parameters in Ref.~\cite{Wang:2011km} and depict \lcdm along with RDE model for different values of the DE sound speed $\cs$. The latter has a relevant effect in both CMB angular power spectrum and matter power spectrum (not shown in Ref.~\cite{Wang:2011km}). 

\section{Data and Methodology} \label{Section:data-methodology}

In order to compute cosmological constraints for the cosmological model including DE as given by the DE energy density \eqref{eq:GO-HDE}, we carried out the analysis in two parts. Firstly, we performed an analysis only taking into consideration background data. Secondly, we add data from the CMB anisotropies and Redshift-Space-Distortions (\texttt{RSD}) that constrain the model further. 

Data constraining the background evolution of the model include: Baryon Acoustic Oscillations (\texttt{BAO}) from Refs.~\cite{BOSS:2016wmc,2011,Ross:2014qpa}, Pantheon supernovae (\texttt{SNe}) data set from Ref.~\cite{Pan-STARRS1:2017jku}, and the SH0ES local measurement of the Hubble constant (\texttt{H0}) from Ref.~\cite{Riess:2021jrx} that we introduce as a Gaussian prior. As for data constraining linear order perturbations   we take in information from CMB lensing (\texttt{lensing}) as well as temperature and polarisation anisotropies of the CMB (\texttt{TTTEEE}) measured by the Planck Collaboration~\cite{Planck:2018vyg}, and a compilation of \texttt{RSD} as explained in Ref.~\cite{Arjona:2020yum}.

As discussed previously, we implemented the HDE cosmological model by considering a DE fluid with DE equation of state \eqref{eq:EOS_GO}, constant sound speed $\cs$, and vanishing anisotropic stress. We carried out the implementation in the widely used Boltzmann solver \texttt{CLASS}. For a given set of cosmological parameters, the code computes all the relevant quantities (e.g., luminosity distances, CMB angular power spectrum) so that theoretical predictions can be compared with astrophysical measurements. We performed a Markov Chain Monte Carlo (MCMC) statistical analysis by sampling the parameter space with the code \texttt{Monte Python}~\cite{Audren:2012wb,Brinckmann:2018cvx}. The latter is linked to \texttt{CLASS} and samples the parameter space with the default Metropolis-Hastings algorithm. In a first stage, a covariance matrix is adjusted so that the acceptance rate is $\approx 0.25$. Then, in a second stage of the analysis, the covariance matrix is fixed and the code performs $\sim 10^6$ iterations until reaching convergence which we estimate with the Gelman-Rubin statistic $R$ satisfying the condition $R-1 \lesssim 0.01$ for all the varying parameters. We marginalise over the following cosmological parameters: baryon density today $\omega_{\mathrm{b}} \equiv \Omega_{\mathrm{b}} h^2$; cold dark matter density today $\omega_{\mathrm{cdm}} \equiv \Omega_{\mathrm{cdm}} h^2$; $100\times$ angular size of sound horizon at redshift $z_{\star}$ (redshift for which the optical depth equals unity) $100\theta_{\star}$; Log power of the primordial curvature perturbations $\ln 10^{10}A_{\rm{s}}$; scalar spectrum power-law index $n_{\rm{s}}$; Thomson scattering optical depth due to reionisation $\tau$; sound speed squared on the rest-frame of the fluid $\log \cs$; parameters determining the holographic DE density $\alpha$ and $\beta$. In our MCMC analyses we also marginalise over a few nuisance parameters whose number depends on the specific probe combination. For common cosmological parameters we use the same prior range as specified in Table 1 of Ref.~\cite{Planck:2013pxb}.

When performing the first part of our statistical analysis (only background data), we vary the parameters $\omega_{\mathrm{b}}$, $\omega_{\mathrm{cdm}}$, $\alpha$, and $\beta$ (and $H_0$ when introduced as a Gaussian prior); the HDE parameters $\alpha$ and $\beta$ were introduced with an unbounded flat prior. In the second part of our analysis (taking into consideration background data as well as CMB anisotropies and Redshift-Space-Distortions), we set $\beta=\alpha/2$ as argued earlier in Section~\ref{Section:model}. Therefore, we vary the parameters  $\omega_{\mathrm{b}}$,  $\omega_{\mathrm{cdm}}$, $100\theta_{\star}$, $\ln 10^{10}A_{\rm{s}}$, $n_{\rm{s}}$, $\tau$, $\log \cs$, and $\alpha$. In this case for the parameters describing the DE fluid we use the prior range specified in Table~\ref{tab:hresult}.

While the analysis carried out in Ref.~\cite{Wang:2011km} seems to have fixed the DE sound speed $\cs=1$, here we marginalise over $\cs$. In Figure~\ref{fig_PRD_Observables_best} we show CMB angular power spectra along with matter power spectra for different values of $\cs$. It becomes clear that $\cs$ plays a part in the analysis, hence fixing the DE sound speed might lead to biased constraints.    
\begin{table}[http]
\centering 
\begin{tabular}{c c} 
\hline\hline 
Parameter & Prior range \\ 
\hline 
$\alpha$ & [$0.01$, 1.35$]$ \\
$\log \cs$ & [$-10$, $0$]\\
\hline 
\end{tabular}
\caption{Flat prior bounds used in the full analysis including background and linear perturbations.}
\label{tab:hresult}
\end{table}

\section{Results and discussion}
\label{section:results}

\begin{figure}[http]
\centering
\includegraphics[scale=1.]{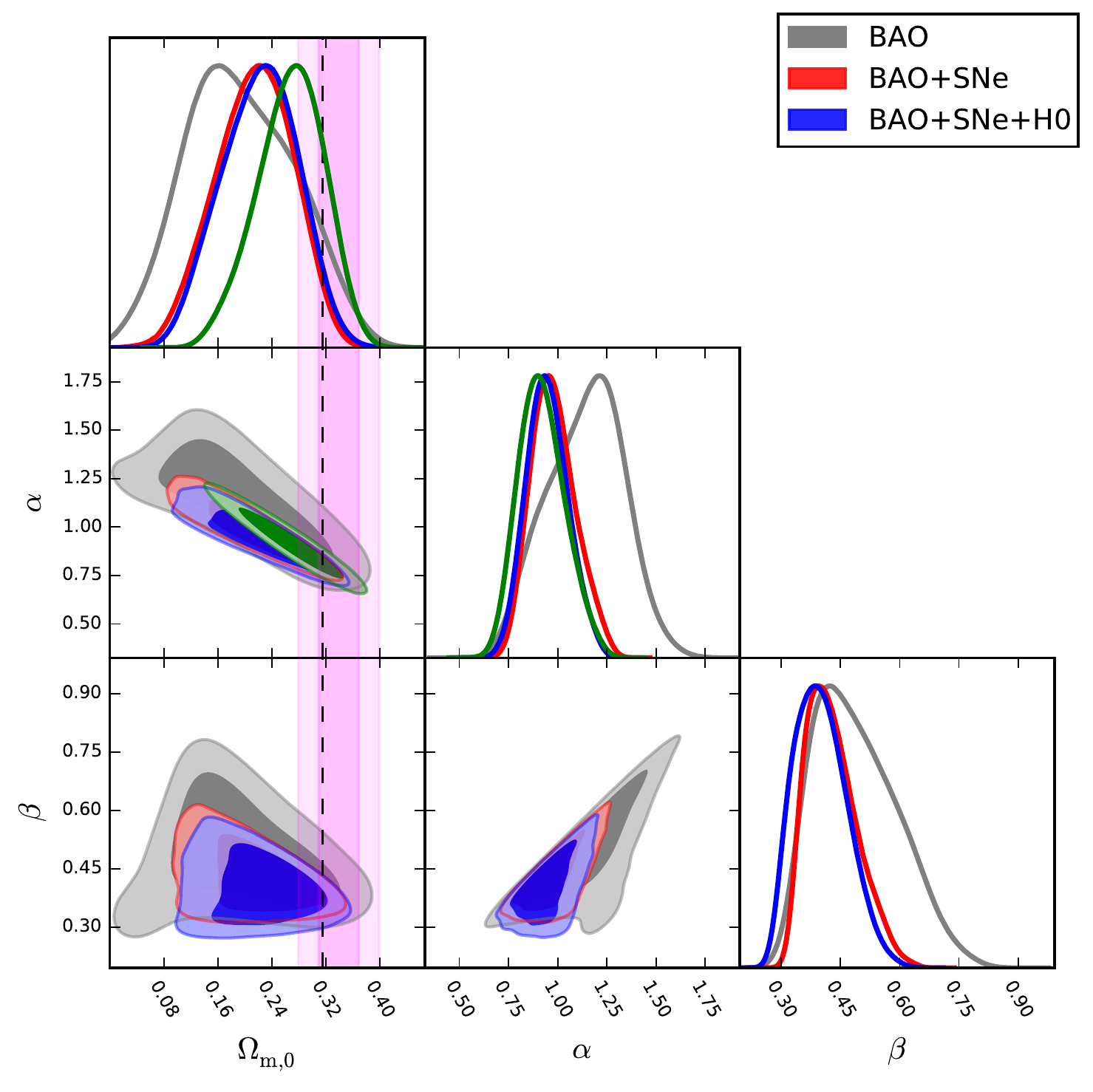}
\caption{1D marginalised likelihoods as well as confidence contours (i.e., $68\%$ and $95\%$) for the HDE (i.e., gray, red, and blue) cosmological model. Green contours and curves show results for the RDE model with \texttt{BAO+SNe+H0}. Note that here we plot the matter density parameter $\Omega_{\mathrm{m},0} \equiv (\omega_{\mathrm{b}}+\omega_{\mathrm{cdm}})/h^2$ which is a derived parameter in our analysis. Dashed, vertical line indicates the result obtained by the Planck Collaboration using the standard cosmological model (see Table 2, column \texttt{TTTEEE+lowE+lensing} in Ref.~\cite{Planck:2018vyg}). Vertical bands indicate DES results ($68\%$ and $95\%$ confidence intervals) for \lcdm reported in Ref.~\cite{DES:2021wwk}. }
\label{fig:constraints-background-alone}
\end{figure}

\begin{figure}[http]
\centering 
\includegraphics[scale=0.74]{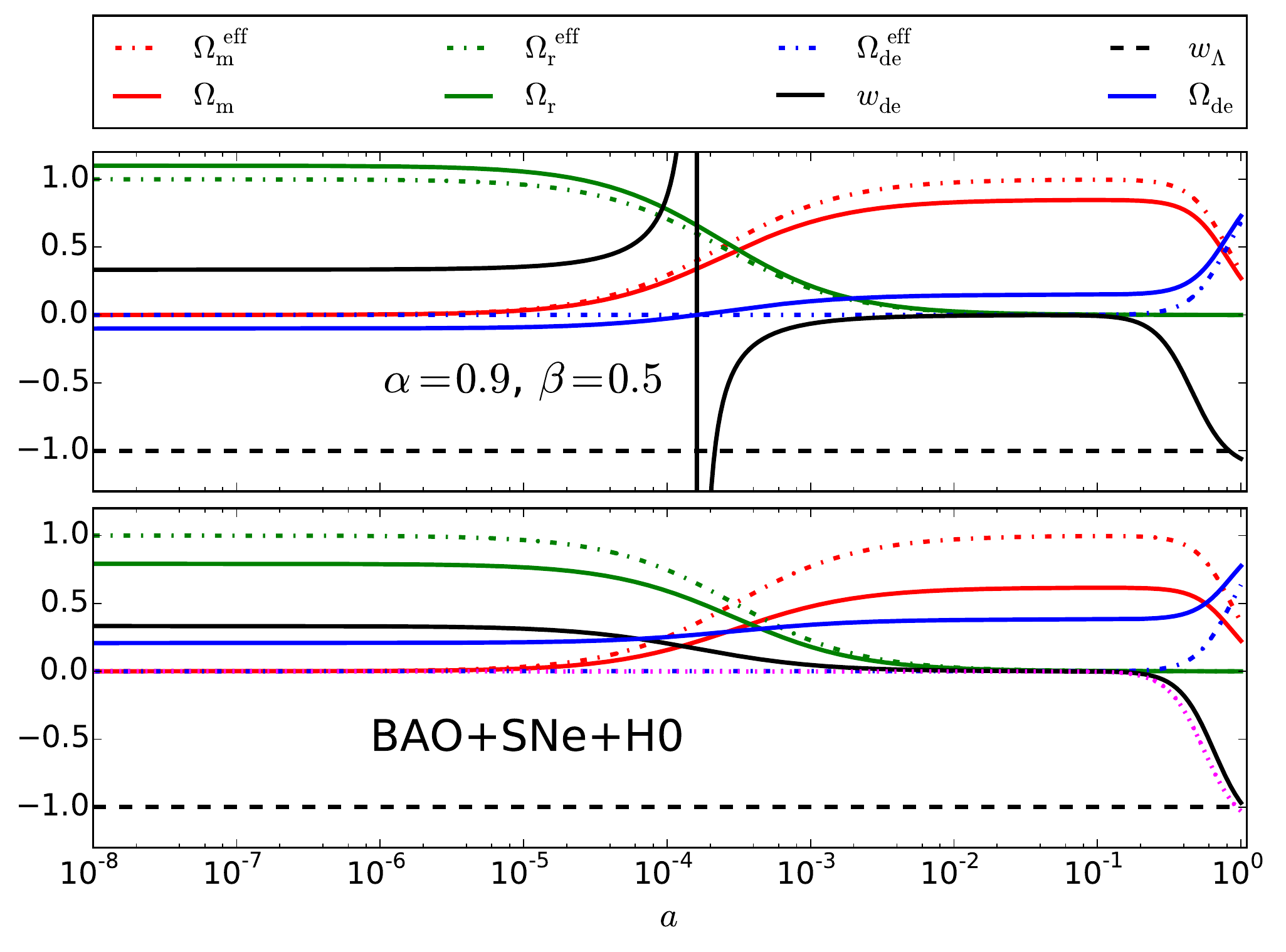}
\caption{Evolution of parameter densities and HDE equation of state $\wde$ for the HDE model. In the lower panel we use the best fit values for \texttt{HDE:BAO+SNe+H0} ($\alpha=0.906$, $\beta=0.349$, $H_0=72.90\,\rm{km}\,\rm{s}^{-1}\,\rm{Mpc}^{-1}$, $\Omega_{\rm{m},0}=0.224$). The magenta, dotted curve shows $\wde$ for the best fit of \texttt{RDE:BAO+SNe+H0} ($\alpha=0.893$, $H_0=73.12\,\rm{km}\,\rm{s}^{-1}\,\rm{Mpc}^{-1}$, $\Omega_{\rm{m},0}=0.281$). We set $\Omega_{\rm{r},0} = 8$.$5 \times 10^{-5}$, and for the upper panel we use  $\Omega_{\rm{m},0} = 0$.$2713$.} 
\label{fig:DE_Cons}
\end{figure}

MCMC results for the first part of our analysis constraining the background evolution are summarised in Fig. \ref{fig:constraints-background-alone} and Table \ref{Table:constraints-background-alone}. Mean values for the HDE parameters $\alpha$ and $\beta$ are in good agreement with previous works using different data sets \cite{PhysRevD.81.083523,Akhlaghi:2018knk,oliveros2022barrow}. Although we do not put any hard bound for $\alpha$ and $\beta$ in our MCMC analysis, in Fig. \ref{fig:constraints-background-alone} we can clearly see that there are no samples in the region satisfying $\alpha<2\beta$. This is due to the fact that in this region the HDE density becomes negative and we have required the condition $\rhode \geq 0$ to hold. Figure~\ref{fig:DE_Cons} shows the behaviour of parameter densities and $\wde$ for a model having $\alpha<2\beta$ (upper panel) as well as the best fit for the case \texttt{HDE:BAO+SNe+H0} in Table~\ref{Table:constraints-background-alone} (lower panel). While in the upper panel we clearly see that $\wde$ has a singularity when $\Omega_{\mathrm{de}}$ changes sign, the lower panel shows a non-negligible amount of HDE during both radiation dominated epoch and DM domination. This behaviour can be understood if we note that, whatever probe combination in Table~\ref{Table:constraints-background-alone}, samples for $\alpha$ and $\beta$ satisfy $\alpha>2\beta$ and therefore HDE effectively contributes to matter and radiation [see Eqs. \eqref{eq:omega-radiation-effective-today}-\eqref{eq:omega-matter-effective-today}]. Consequently, we observe in Fig. \ref{fig:constraints-background-alone} a degeneracy between the matter parameter density $\Omega_{\mathrm{m},0}$ and the HDE parameter $\alpha$. The degeneracy is even more evident for the green contours showing results for the RDE model.

\begin{table*}[http]
\centering
\begin{tabular}{c c c c c}
\hline
Parameter  & \texttt{HDE:BAO} & \texttt{HDE:BAO+SNe} & \texttt{HDE:BAO+SNe+H0} & \texttt{RDE:BAO+SNe+H0} \\
\hline
$\omega_{\mathrm{b}}$ & $0.0195^{-0.0144}_{+0.0051}$ & $0.0151^{-0.0097}_{+0.0025}$ & $0.0176^{-0.0108}_{+0.0039}$ & $0.0228^{-0.0106}_{+0.0048}$\\
$\omega_{\mathrm{cdm}}$  & $0.0688^{-0.0585}_{+0.0334}$ & $0.0821^{-0.0261}_{+0.0363}$ & $0.1010^{-0.0319}_{+0.0384}$ & $0.1202^{-0.0292}_{+0.0408}$\\
$\beta $  & $0.49^{-0.13}_{+0.08}$ & $0.43^{-0.08}_{+0.04}$ & $0.40^{-0.08}_{+0.06}$ & --\\
$\alpha$  & $1.14^{-0.18}_{+0.22}$ & $0.98^{-0.13}_{+0.10} $ & $ 0.95^{-0.11}_{+0.10} $ & $0.92^{-0.13}_{+0.10}$ \\
$\Omega_{\rm{m},0}$ & $0.195^{-0.091}_{+0.077}$ & $0.214^{-0.054}_{+0.063}$ & $0.222^{-0.054}_{+0.062}$ & $0.268^{-0.045}_{+0.056}$ \\
\hline
\end{tabular}
\caption{Mean values and $68\%$ confidence limits on cosmological parameters. Here we only use data constraining the background.}
\label{Table:constraints-background-alone}
\end{table*}

\begin{figure}[http]
\centering
\includegraphics[scale=1.05]{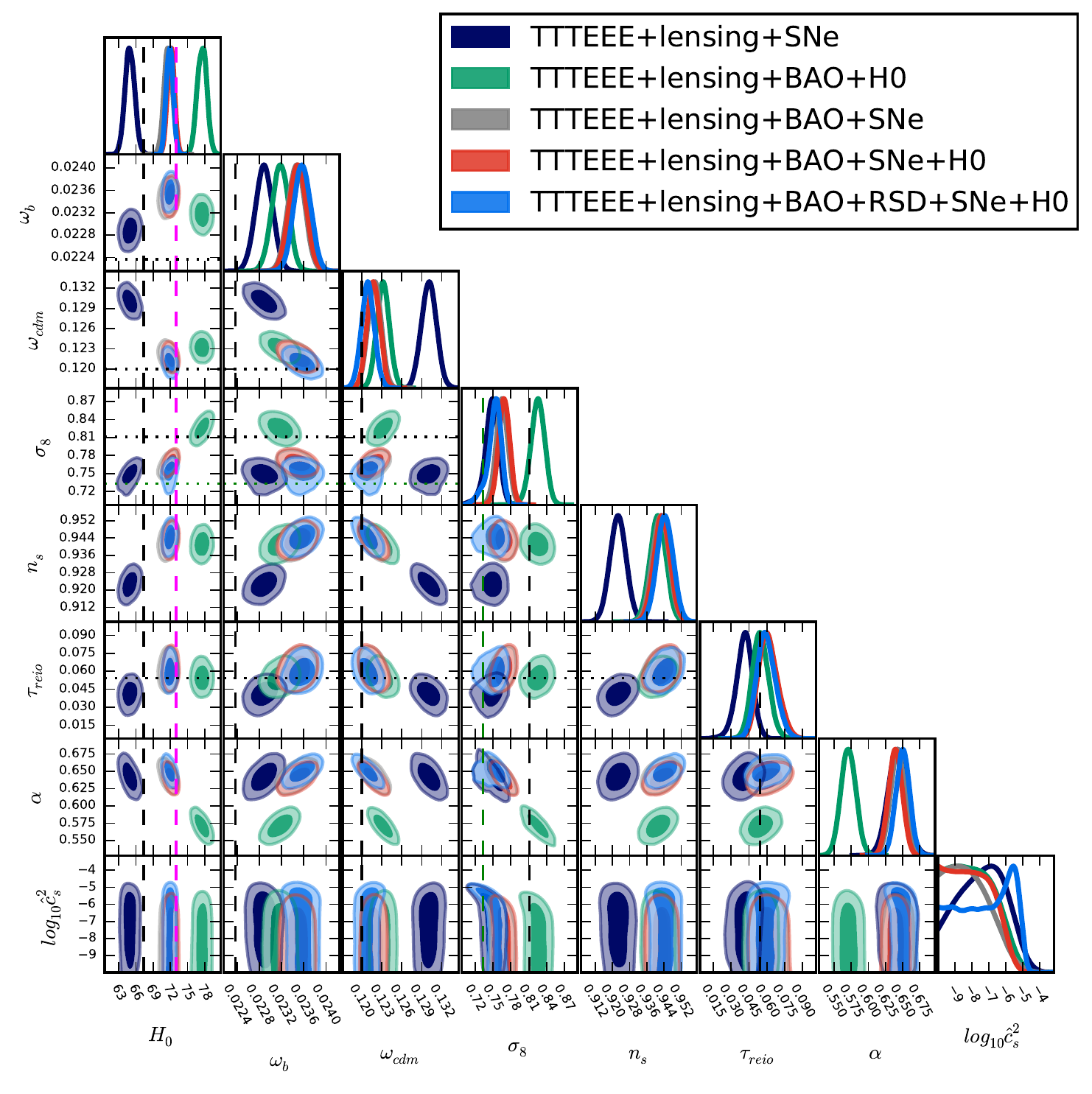}
\caption{1D marginalised likelihoods as well as confidence contours (i.e., $68\%$ and $95\%$) for the RDE cosmological model. Dashed, vertical and dotted, horizontal black lines indicate the results obtained by the Planck Collaboration using the standard cosmological model (see Table 2, column \texttt{TTTEEE+lowE+lensing} in Ref.~\cite{Planck:2018vyg}). Note that here we plot the Hubble constant $H_0$ and the strength of matter clustering $\sigma_8$ which are derived parameters in our analysis. Vertical, dashed, magenta line indicates SH$0$ES value $73.04\,\rm{km}\,\rm{s}^{-1}\,\rm{Mpc}^{-1}$ \cite{Riess:2021jrx}. Dashed, vertical and dotted, horizontal green lines indicate DES value $\sigma_8 = 0.733$ for the analysis of large scale structure combining three two-point correlation functions \cite{DES:2021wwk}.}
\label{fig:constraints}
\end{figure}

\begin{figure}[http]
\centering 
\includegraphics[scale=0.74]{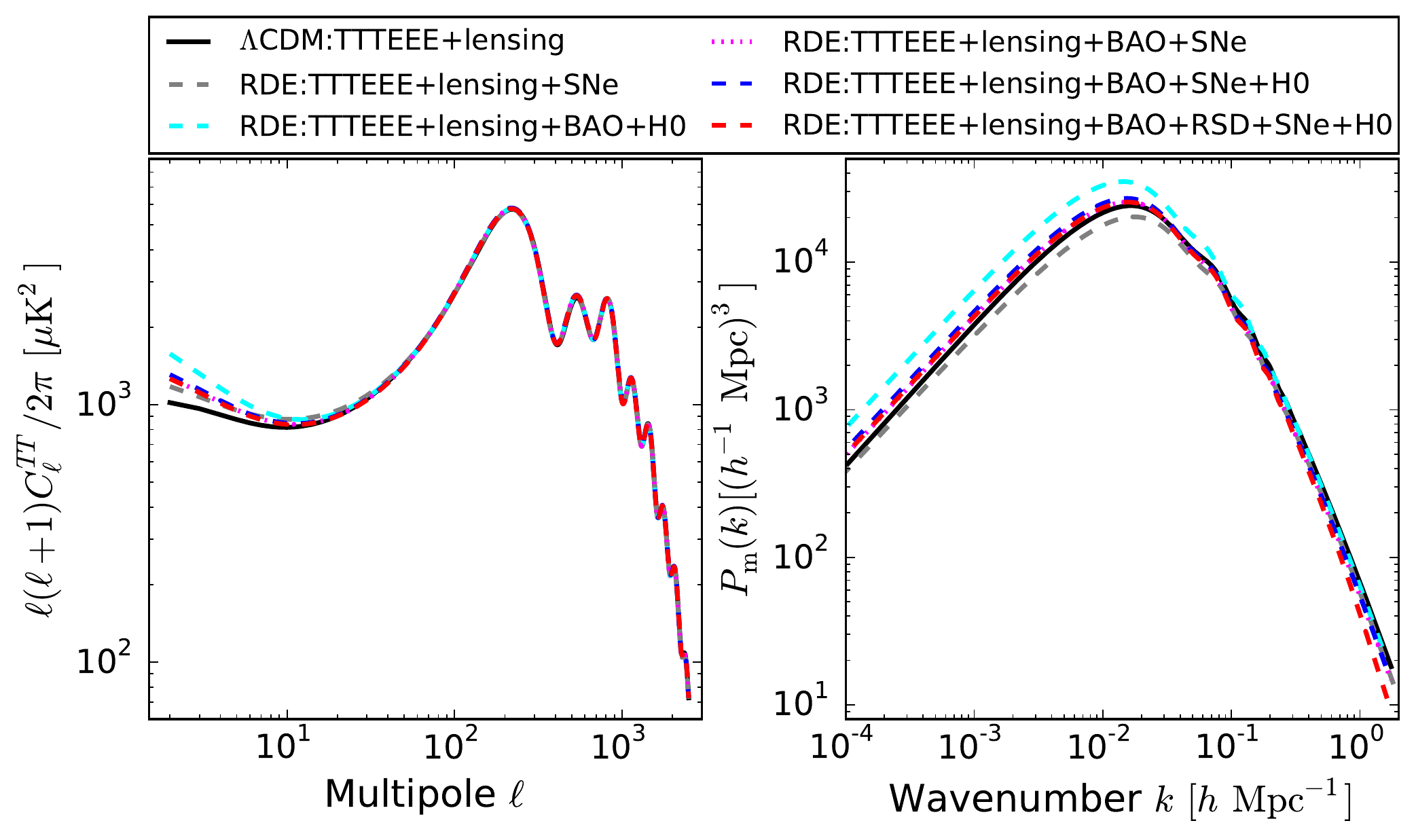}
\caption{Left: CMB temperature angular power spectrum $C_{l}^{TT}$. Right: linear theory matter power spectrum $P_{\rm{m}}(k,z=0)$. We plot the baseline result reported by the Planck Collaboration ($\Lambda$CDM) along with best fits of analyses in Fig.~\ref{fig:constraints} for the RDE model. Our baseline result (\texttt{TTTEEE+lensing+BAO+RSD+SNe+H0}) has the following best fit values for cosmological parameters: $\omegab = 0.02351$, $\omegacdm = 0.1214$, $100\theta_{\rm{s}} = 1.04087$, $\As = 3.014$, $\ns = 0.9440$, $\tau = 0.0551$, $\alpha = 0.651$, $\log \cs = -6$, $\sigma_8 = 0.736$, $H_0 = 71.44\,\rm{km}\,\rm{s}^{-1}\,\rm{Mpc}^{-1}$.}
\label{figObservables_best-1}
\end{figure}

\begin{figure}[http]
\centering 
\includegraphics[scale=0.86]{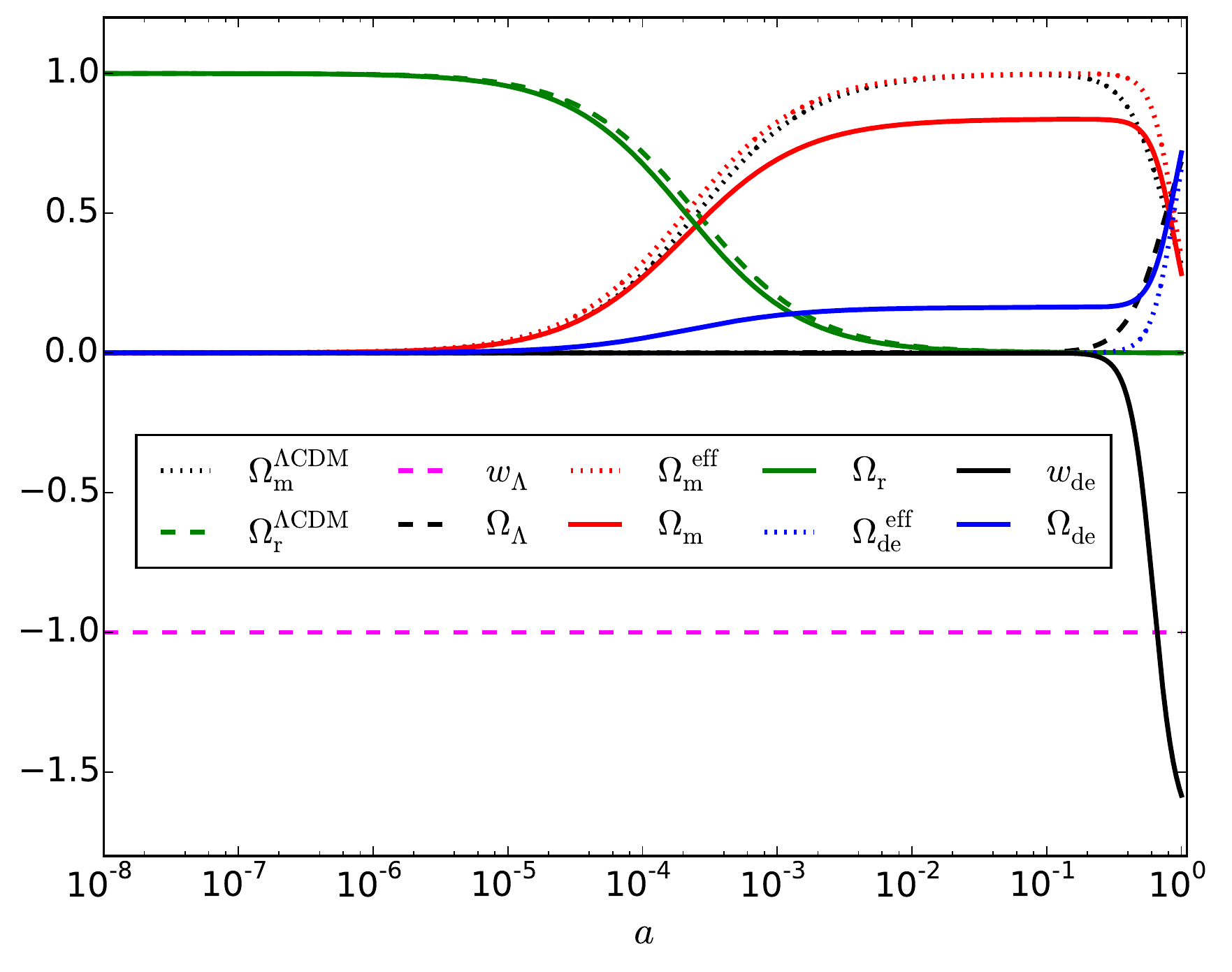}
\caption{Evolution of parameter densities and DE equation of state $\wde$ for the RDE model. Here we use the best fit values for our baseline result (\texttt{TTTEEE+lensing+BAO+RSD+SNe+H0}) analysis in Fig.~\ref{fig:constraints}. For sake of comparison we also plot the \lcdm baseline result by the Planck Collaboration.}
\label{figObservables_best-2}
\end{figure}

Next we explain the second part of our analysis where data constraining linear order perturbations are also considered and we set  $\beta=\alpha/2$. Statistical information derived from our MCMC results is presented in Table~\ref{Table:constraints-background-perturbations}, while Fig.~\ref{fig:constraints} contains confidence contours and marginalised 1D posteriors. Several comments can be made. Firstly, by comparing Figs.~\ref{fig:constraints-background-alone} and \ref{fig:constraints} we note that while background data sets seem compatible with each other, also taking Planck data into consideration introduces a discordance in the determination of some cosmological parameters (see dark, blue and light, green contours in Fig.~\ref{fig:constraints}), namely, $H_0$, $\omega_{\mathrm{cdm}}$, $\sigma_8$, $n_\mathrm{s}$, and $\alpha$. Secondly, with regard to the RDE parameter $\alpha$, we can see that in this case it is well constrained and its mean value is significantly shifted towards lower values than reported in Table \ref{Table:constraints-background-alone}. Although we use different data sets, constraints for $\alpha$ agree at the $2\sigma$ level with results in Ref.~\cite{Wang:2011km}; there are however noticeable differences in other cosmological parameters such as $\omega_{\rm{b}}$, $\omega_{\rm{cdm}}$, $n_{\rm{s}}$, $\tau$, and $\Omega_{\rm{m},0}$. Thirdly, even though the sound speed squared $\log \cs$ hits the lower bound in the prior and we can only set an upper limit, it becomes clear from our results that $\cs = 1$ is excluded by more than $3\sigma$. This result calls in question the assumption of fixing $\cs$ in Ref.~\cite{Wang:2011km} (presumably to $\cs=1$). A comparison between Figs.~\ref{fig_PRD_Observables_best} and~\ref{figObservables_best-1} clearly confirms that the DE sound speed plays a role in the analysis of RDE. A lower DE sound speed along with changes in $\omega_{\rm{b}}$, $\omega_{\rm{cdm}}$, $n_{\rm{s}}$, and $\tau$ show a much better fit than previous results fixing $\cs=1$. Fourthly, regardless of the probe combination, the only parameter which shows relatively good agreement with the baseline result reported by the Planck Collaboration for the $\Lambda$CDM model is the Thomson scattering optical depth due to reionisation $\tau$. All other parameters in common with the standard cosmological model appear relatively discrepant. Fifthly, our analysis discloses a slight degeneracy between $\alpha$ and the parameters $H_0$, $\omega_{\mathrm{b}}$, $\omega_{\mathrm{cdm}}$, and $\sigma_8$ (see Fig.~\ref{fig:constraints}). Sixthly, while for the analysis in Table \ref{Table:constraints-background-alone} only including background data the DE equation of state $\wde(a=1)\approx -1$ (see lower panel in Fig.~\ref{fig:DE_Cons}), the analysis in Table \ref{Table:constraints-background-perturbations} also including CMB and RSD data yields $\wde(a=1) < -1$ (see Fig.~\ref{figObservables_best-2}). Then, despite having a present DE budget similar to the \lcdm ($\Omega_{\mathrm{de}}(a=1)\approx 0.7$), we conclude the RDE model struggles to simultaneously fit low and high redshift data.

\begin{table*}[http]
\centering
\resizebox{\columnwidth}{!}{%
\begin{tabular}{c c c c c c}
\hline
Parameter  & \texttt{TTTEEE+lensing+SNe} & $\left\lbrace\dots\right\rbrace$\texttt{+BAO} & $\left\lbrace\dots\right\rbrace$\texttt{+H0} & $\left\lbrace\dots\right\rbrace$\texttt{+RSD} & \texttt{TTTEEE+lensing+BAO+H0} \\
\hline
$\omega_\mathrm{b} $ & $0.02288^{-0.00015}_{+0.00016}$ & $0.02347\pm 0.00015$ & $0.02349\pm 0.00015$ & $0.02357\pm 0.00015$ & $0.02317\pm 0.00015$\\
$\omega_{\mathrm{cdm}}$  & $0.1301\pm 0.0011$ & $0.1219\pm 0.0010$ & $0.1218\pm 0.0010$ & $0.1209\pm 0.0010$ & $0.1232^{-0.0010}_{+0.0009}$ \\
$H_0$ &  $64.95^{-0.83}_{+0.84}$ & $71.65^{-0.73}_{+0.75} $ & $72.12^{-0.62}_{+0.60}$ & $71.91^{-0.61}_{+0.59}$ & $77.55^{-0.85}_{+0.82}$ \\
$ \sigma_8$  & $0.748^{-0.010}_{+0.013} $ & $0.765\pm 0.010$ & $0.769\pm 0.009$ & $0.752^{-0.008}_{+0.014}$ & $0.825\pm 0.012$\\
$n_{\rm{s}}$ &  $0.9226\pm 0.0038$ & $0.9425\pm 0.0038$ & $0.9431\pm 0.0037$ & $0.9445^{-0.0037}_{+0.0038}$ & $0.9412\pm 0.0037$ \\
$\tau $  & $0.0411^{-0.0061}_{+0.0071}$ & $0.0605^{-0.0085}_{+0.0072}$ & $0.0612^{-0.0090}_{+0.0070}$ & $0.0596^{-0.0086}_{+0.0070}$ & $0.0547^{-0.0073}_{+0.0065}$ \\
$\log \cs $  & $-7^{-2}_{+1}$ & $<-8$ & $<-7$ & $<-6$ & $ -8^{-2}_{+1}$ \\
$\alpha$  & $0.642\pm 0.013$ & $0.643\pm 0.011 $ & $0.640\pm 0.010$ & $0.650\pm 0.010$ & $ 0.571\pm 0.011 $  \\
$\Omega_{\rm{m},0}$ & $0.363\pm 0.011$ & $0.283^{-0.007}_{+0.006}$ & $0.279^{-0.006}_{+0.005}$ & $0.279\pm 0.005$ & $0.243^{-0.005}_{+0.006}$  \\
$100\theta_{\star}$ & $1.03999\pm 0.00028$ & $1.04084\pm 0.00028$ & $1.04088\pm 0.00028$ & $1.04092\pm 0.00028$ & $1.04076\pm 0.00028$ \\
$S_8$ & $0.822^{-0.012}_{+0.014}$ & $0.744\pm 0.010$ & $0.742\pm 0.010$ & $0.726^{-0.010}_{+0.014}$ & $0.743\pm 0.010$ \\
\hline
\end{tabular}
}
\caption{Mean values and $68\%$ confidence limits on cosmological parameters for the RDE model. Here $\left\lbrace\dots\right\rbrace$ stands for the inclusion of data from column on the left.}
\label{Table:constraints-background-perturbations}
\end{table*}

\begin{figure}
\centering
\includegraphics[scale=0.74]{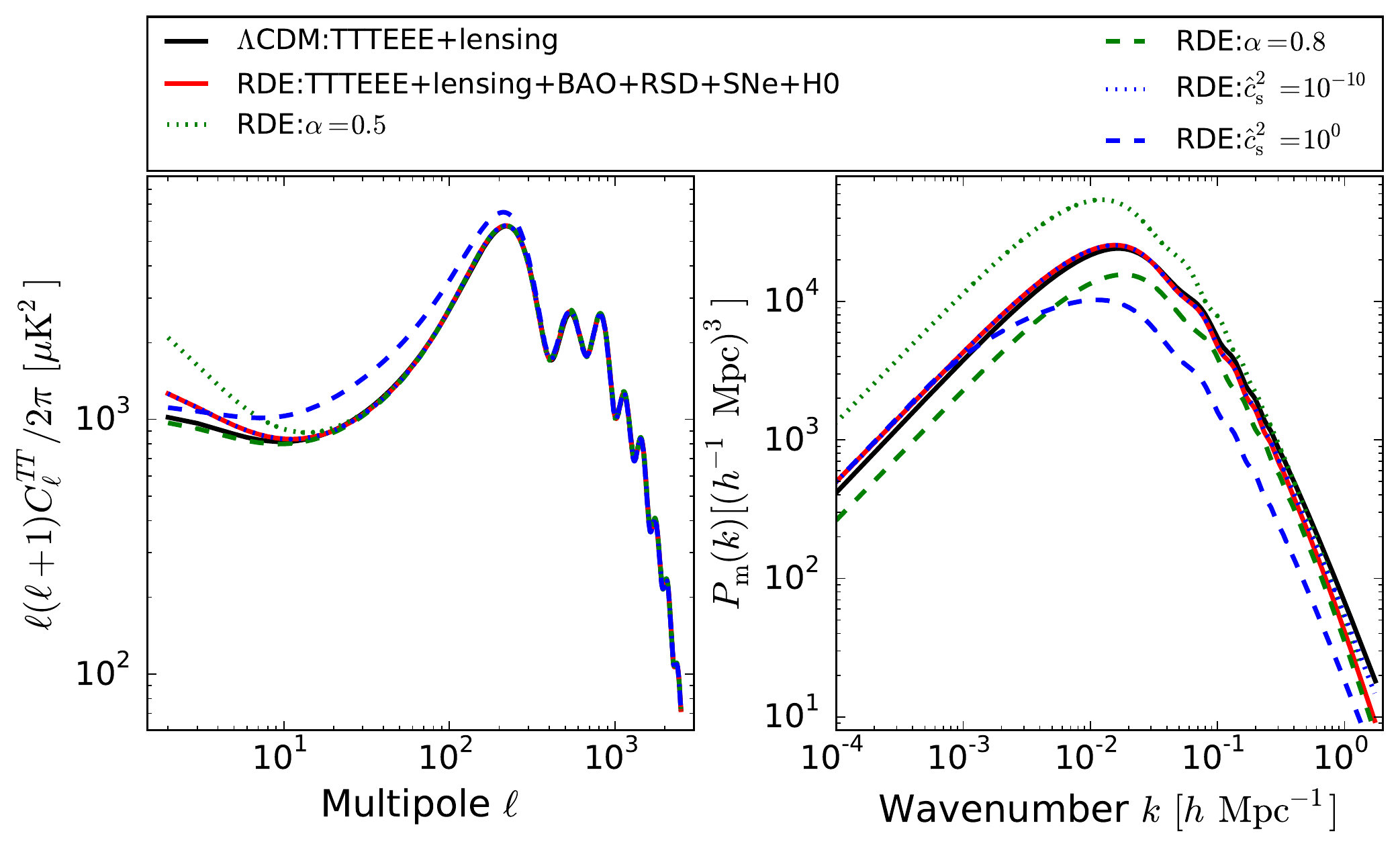}
\caption{Left: CMB  angular power spectrum. Right: linear theory matter power spectrum. Solid, black lines show the Planck Collaboration baseline result for the standard model. Solid, red lines show our baseline result for the RDE model. Green, dotted (dashed) line shows the effect of decreasing (increasing) $\alpha$ with respect to the best fit value. Blue, dotted (dashed) line shows the effect of decreasing (increasing) $\cs$ with respect to the best fit value.}
\label{fig:alpha-cs2-cmb-pk}
\end{figure}

In Fig.~\ref{figObservables_best-1}  we show the CMB TT angular power spectrum and the matter power spectrum corresponding to the best fits of analyses in Table \ref{Table:constraints-background-perturbations} and Fig.~\ref{fig:constraints}. For the sake of comparison we also depicted the Plack baseline result for the \lcdm model. Main differences in the CMB angular power spectrum appear on very large angular scales where cosmic variance dominates the error budget and the Sachs-Wolfe effect becomes important: the enhancement of power at small $\ell$ is due to lower values for the spectral index than in \lcdm as well as the evolution of gravitational potentials (affected by $\alpha$ and $\cs$) when DE dominates the energy budget (see also left panel of Fig.~\ref{fig:alpha-cs2-cmb-pk}). Concerning the matter power spectrum, we can see that it is heavily modified with respect to the \lcdm solution depending on the data set used in the analysis. Except for the combination \texttt{TTTEEE+lensing+SNe}, data favour more power on large scales than in the \lcdm model. Except for the combination \texttt{TTTEEE+lensing+BAO+H0}, data favour less power on small scales than in the standard model. Our baseline result \texttt{TTTEEE+lensing+BAO+RSD+SNe+H0} keeps the angular acoustic scale in good agreement with the Planck Collaboration baseline result. Right panel of Fig.~\ref{fig:alpha-cs2-cmb-pk} shows the effect of changing $\alpha$ and $\cs$ that we explain below.  

In what follows we focus on our baseline result \texttt{TTTEEE+lensing+BAO+RSD+SNe+H0} which seems to bring in relative good agreement CMB and Large Scale Structure data. First, note that although we do not use data from the Dark Energy Survey (DES), our derived constraints for $S_8$ and $\Omega_{\rm{m},0}$ (see Table~\ref{Table:constraints-background-perturbations}) are compatible with DES measurements are $S_8=0.776\pm 0.017$ and $\Omega_{\rm{m},0}=0.339^{+0.032}_{-0.031}$ within $\approx2\sigma$. Second, we can see in the right panel of Fig.~\ref{figObservables_best-1} than on large scales (wavenumber $k\lesssim 10^{-2}\,\rm{h}\,\rm{Mpc}^{-1}$) beyond the reach of current galaxy surveys, the RDE model predicts more power than the standard model. This difference is due to a few changes with respect to the concordance model: i) a smaller spectral index ($n_{\rm{s}}=0.9649$ in $\Lambda$CDM); ii) a slight shift in the pivot scale; iii) differences in the evolution of matter perturbations in the RDE model (see Fig.~\ref{fig:matter-perturbations}). Third, we constrain the redshift dependence in the linear matter power spectrum by using the parameter $f\sigma_8$ through the likelihood \texttt{RSD} which relies on linear perturbation theory.\footnote{We also computed non-linear corrections for the RDE using HALOFIT \cite{Smith:2002dz}. However, since HALOFIT is optimised for the standard model, we do not use it in this work. Even when including these non-linear corrections we observe less power on small scales in the RDE model than predicted by the \lcdm model. Non-linear evolution of DE perturbations could also be investigated along the lines explained in Refs. \cite{Abramo:2007iu,Abramo:2008ip}.} Our constraint for the strength of matter clustering $\sigma_8$ turns out to be lower than in the \lcdm model Planck baseline. Since matter velocity perturbations might be greater in RDE than in \lcdm (see Fig.~\ref{fig:matter-perturbations}), we expect less matter clustering in RDE. Moreover, since the rms linear theory mass fluctuation in a sphere of radius $R=8\,\rm{Mpc}\,\rm{h}^{-1}$ at $z=0$
\begin{equation}
(\sigma_8)^2 \equiv \frac{1}{2\pi^2} \int d \log k\,W^2(kR)\,k^3\,P_{\rm{m}}(k),
\label{eq:sigma8}
\end{equation}
where $W(kR)$ is a spherical top-hat filter, is predominantly determined by contributions on small scales and the matter power spectrum $P_{\rm{m}}$ predicts less power in RDE than in $\Lambda$CDM, we can expect $\sigma_8^{\rm{RDE}} < \sigma_8^{\Lambda\rm{CDM}}$ as we indeed found.
  
\begin{figure}[http]
    \centering
    \includegraphics[scale=0.86]{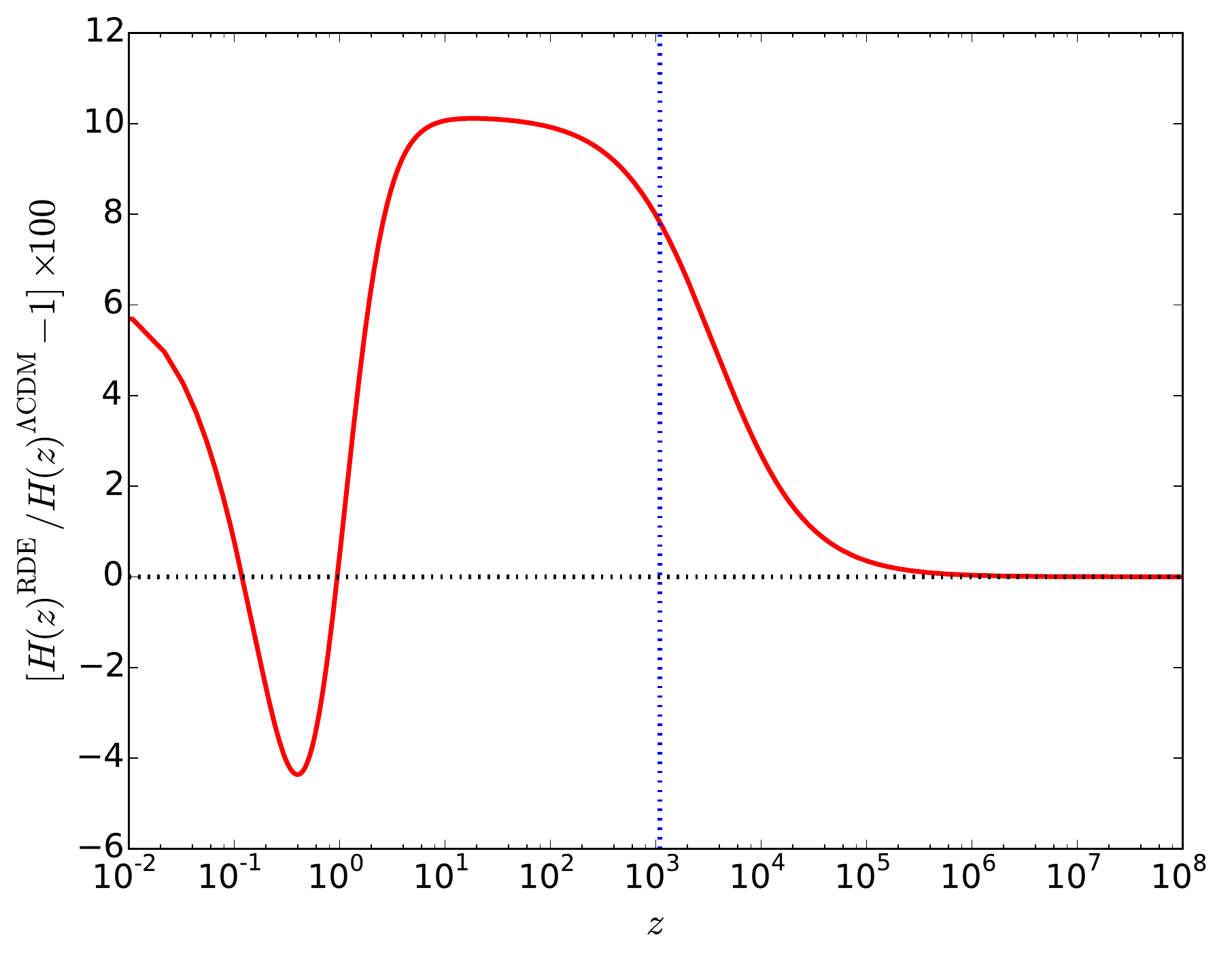}
    \caption{Percentage difference in the Hubble parameter for the RDE model (best fit of our baseline result) with respect to the Planck \lcdm baseline result. Blue, vertical dotted line indicates $z_{\star}$.}
    \label{fig:H-of-z}
\end{figure}

Another interesting aspect of our analysis concerns the constraint for the Hubble constant $H_0$. Fig.~\ref{fig:constraints} indicates that the RDE model can simultaneously relax the current tension in $H_0$ and $\sigma_8$ (see case \texttt{TTTEEE+lensing+BAO+SNe}). The angular acoustic scale 
\begin{equation}
    \theta_{\rm{s}} = \frac{r_{\rm{s}}(z_{\star})}{D_{\rm{A}}(z_{\star})},
    \label{eq:acoustic-scale}
\end{equation}
is pretty well constrained by CMB observations. In Eq.~\eqref{eq:acoustic-scale}, $r_{\rm{s}}$ and $D_{\rm{A}}$ respectively denote the sound horizon
\begin{equation}
    r_{\rm{s}} = \int_{z_{\star}}^{\infty} \frac{c_{\rm{s}}(z)}{H(z)} dz,
\end{equation}
and the comoving angular diameter distance
\begin{equation}
    D_{\rm{A}} = \int_{0}^{z_{\star}} \frac{dz}{H(z)}. 
\end{equation}
While for the \lcdm Planck baseline 
\begin{equation}
100\theta_{\rm{s}}=1.04110,\quad r_{\rm{s}}(z_{\star})=144.531055\,\rm{Mpc},\quad d_{\rm{A}}(z_{\star})=12.738778\,\rm{Mpc},
\end{equation}
our baseline result (the best fit) 
\begin{equation}
100\theta_{\rm{s}}=1.04087,\quad r_{\rm{s}}(z_{\star})=136.800000\,\rm{Mpc},\quad d_{\rm{A}}(z_{\star})=12.053113\,\rm{Mpc}, 
\end{equation} 
where the comoving angular diameter distance is related to the angular diameter distance $d_{\rm{A}}$ via $D_{\rm{A}}(z)=(1+z) d_{\rm{A}}$. The RDE model decreases both the sound horizon and the comoving angular diameter distance while keeping the angular acoustic scale in good agreement with the \lcdm solution. These changes can be understood from Figure~\ref{fig:H-of-z}. On the one hand, in the RDE model the expansion rate is enhanced with respect to the standard model in two stages ($10^{0} \lesssim z \lesssim 10^5$ and at late times $z \lesssim 10^{-1}$). On the other hand, for a relatively short period of time the universe expands faster in \lcdm than in RDE for $ 10^{-1} \lesssim z \lesssim 10^0$. While the Early Dark Energy model of Ref. \cite{PhysRevLett.122.221301} mainly changes the sound horizon through the enhancement of $H(z)$ prior to recombination, the RDE model introduces changes in the expansion rate prior and post recombination.  

Although our RDE baseline result brings into $\approx2\sigma$ agreement $H_0$ and $\sigma_8$, other parameters get shifted in order to maintain the fit to primary CMB and RSD data.\footnote{In the EDE scenario alleviating the $H_0$ tension, a similar situation occurs. However, the EDE exacerbates the $\sigma_8$ tension while including large scale structure data in the analysis \cite{PhysRevD.102.043507}.} Besides the shift of $\ns$ towards lower values than allowed in the \lcdm analysis of Planck, we also obtain a value of baryon matter density $\omegab$ higher than in the Planck \lcdm baseline result. This value actually exacerbates the existing $\approx 2\sigma$ discrepancy in the \lcdm model with values inferred from Big Bang Nucleosynthesis (BBN) \cite{Cooke:2017cwo}.\footnote{Recent analyses using an improved rate of deuterium burning estimate a $\omegab=0.02233\pm 0.00036$ in excellent agreement with the Planck baseline result \cite{Mossa:2020gjc}. This value is discrepant with our baseline RDE result at the $3\sigma$ level.}     

We finalize our discussion by comparing our findings with previous results. First, whereas in Ref.~\cite{Wang:2011km} authors do not seem to have properly included radiation in their analysis when solving for the background evolution (see their Eq. 5 for the Hubble parameter), here we have derived the full expression in Eq.~\eqref{eq:HDE-solution-H-Ricci} [or the more general Eq.~\eqref{eq:HDE-solution-H}]. Second, since the RDE model can cross the phantom divide $\wde=-1$ we have used the PPF formalism so that perturbations behave properly. Authors in Ref.~\cite{Wang:2011km} split the whole region of $\wde$ into three regions ending up with a much more involved implementation. Third,  while in Ref.~\cite{Wang:2011km} values $n_{\rm{s}}>1$ are preferred, our analysis favours values $n_{\rm{s}}<1$. This is due to the much better  constraining power added by Planck data on small scales in comparison to WMAP data. 

\section{Conclusions}
\label{Section:conclusions}

Holographic Dark Energy (HDE) seems to offer a plausible alternative to the simple, but troublesome cosmological constant. While the holographic approach does not require modifications to General Relativity, neither does it require new dynamical Dark Energy (DE) fields, it does need an arbitrary choice. In these kinds of DE models the length scale fixing the maximum energy density in the effective field theory becomes important. In the literature there exist a few ultraviolet/infrared relationships that avoid the problem of causality as well as the coincidence issue. In this work we studied two infrared (IR) cut-offs: the Ricci scalar curvature (also dubbed RDE) and its more general version the Granda-Oliveros (GO) IR cut-off.

Thus far most works studying phenomenological aspects of HDE  models have focused on the background. Here we scrutinised HDE models derived from both GO and RDE cut-offs also considering linear order perturbations. Our investigation refined upon previous works as we summarise below.

Concerning the background evolution. We noted that in previous works radiation is not properly taken into account when solving for the Hubble parameter. We showed this might not be a good approximation: depending on the cut-off choice, new terms might appear playing a part in early stages of the universe expansion. As a result, the holographic energy density using the GO cut-off can be radiation-like, matter-like, or DE-like depending on the component that dominates the energy budget. For the RDE IR cut-off the holographic energy density can only be matter-like or DE-like, since the constraint $\alpha=2\beta$ excludes a radiation-like behaviour (see Fig.~\ref{fig:phenomenology-background-1}). Interestingly, the peculiar behaviour in the holographic energy density changes the expansion rate prior and post-recombination (see Fig.~\ref{fig:H-of-z}). Consequently, the RDE cosmological model can decrease both sound horizon and comoving angular diameter distance while keeping good agreement with CMB measurements of the angular acoustic scale. This feature might help to relax the current discrepancy in the Hubble constant between low and high redshift probes.  

Considering a DE fluid having a background evolution matching the RDE model, we proceeded to the investigation of linear order perturbations and assumed our DE as having a constant sound speed $\cs$ as well as vanishing anisotropic stress. We managed to find analytical, approximate solutions for matter and DE perturbations in the regime of matter dominance and when the gravitational potential can be regarded as constant. Our findings show that matter perturbations behave slightly different with respect to the standard cosmological model \lcdm (see Fig.~\ref{fig:matter-perturbations}). Due to the matter-like behaviour of the holographic density in the RDE model, DE clusters in the same way as matter perturbations when its sound speed vanishes. 

Differences in the matter clustering properties of the RDE model with respect to \lcdm become apparent in the matter power spectrum. We implemented the RDE model in the popular Boltzmann solver \texttt{CLASS} so that predictions for the statistical properties of CMB and matter fluctuations could be computed. Whereas previous works artificially considered the possibility of $\wde$ crossing the phantom divide, here we used the PPF formalism fully granting energy-momentum conservation. Another important difference with respect to previous works concerns the treatment we gave to  $\cs$. In the literature, a RDE model with fixed $\cs=1$ has been investigated. Nevertheless, we showed this choice might not be appropriate as it heavily affects matter clustering through changes in the gravitational potential during matter dominance (see Fig.~\ref{fig:gravitational-potential}). 

We computed cosmological constraints for the RDE model marginalising over $\cs$. In our investigation we regarded CMB data (i.e., temperature, polarisation, lensing), baryon acoustic oscillations, supernovae, redshift space distortions, and the local measurement of the Hubble constant. For our baseline result using all data sets we found that $\cs=1$ is excluded at $\gtrsim 3 \sigma$. We also obtained a value of the strength of matter clustering $\sigma_8$ lower than in the \lcdm model and in good agreement ($\approx2\sigma$) with recent DES results. Our constraint on the Hubble constant $H_0$ value is also in good concordance ($\approx2\sigma$) with local, model independent measurements using Cepheid variables. This can be explained by the presence of a turning point in the Hubble parameter $H(z)$ for $z<0.1$ which is still consistent with \texttt{SNe} observations ~\cite{Colgain:2021beg}. While the scalar spectral index $\ns$ takes on lower values in the RDE model than in the \lcdm analysis, the constraint on the baryon density $\omegab$ in our RDE baseline result appears to be in $\approx 3\sigma$ tension with BBN measurements (see Fig.~\ref{fig:constraints} and Table \ref{Table:constraints-background-perturbations}).     

The RDE model has an evolving DE equation of state which according to our baseline result $\wde(z=0)<-1$ (see Fig.~\ref{figObservables_best-2}). However, when only considering background data we found $\wde(z=0)\approx -1$ (see Fig.~\ref{fig:DE_Cons}), thus showing RDE model struggles to simultaneously fit background and perturbations data. It remains to be seen whether or not a more general IR cut-off such as GO having $\alpha>2\beta$ (RDE has $\alpha=2\beta$) might be able to provide a better fit to the background data than RDE while also relaxing tensions in cosmological parameters (e.g., $H_0$, $\sigma_8$, $\omegab$). Our work shows that dynamical DE also having non standard clustering properties may play a part in the solution of discrepancies in cosmological parameters (other possibilities relying on new physics discussed, for instance, in Refs.~\cite{Heisenberg:2022gqk,Clark:2021hlo,Bansal:2021dfh,Rashkovetskyi:2021rwg,Anchordoqui:2021gji,Gomez-Valent:2021cbe,Murgia:2016ccp}). Scalar-Vector-Tensor theories provide a fairly general framework where $\wde(z)$, $\cs(z,k)$, and $\sigma(z,k)$ worthwhile an investigation in light of current and upcoming experiments \cite{Sabla:2022xzj,Cardona:2022lcz}. 

\section*{Acknowledgements}

We are grateful to Giovanni Marozzi and Alexander Oliveros for helpful discussions. WC thanks University of Pisa for hospitality. WC acknowledges financial support from the S\~{a}o Paulo Research Foundation (FAPESP) through grant \#2021/10290-2. This research was supported by resources supplied by the Center for Scientific Computing (NCC/GridUNESP) of the S\~{a}o Paulo State University (UNESP) and the Datacenter CIBioFi (Universidad del Valle). The statistical  analyses  as  well  as  the  plots  were  made with the Python package GetDist \url{https://github.com/cmbant/getdist}.

\section*{Numerical codes}

Modified \texttt{CLASS} code reproducing results in this work can be found in the GitHub branch \texttt{HDE} of the repository  \href{https://github.com/wilmarcardonac/EFCLASS.git}{\texttt{EFCLASS}}.   



\bibliographystyle{JHEP}
\bibliography{paper}

\providecommand{\href}[2]{#2}\begingroup\raggedright\begin{thebibliography}{10}

\bibitem{Planck:2018vyg}
{\scshape Planck} collaboration, \emph{{Planck 2018 results. VI. Cosmological
  parameters}},
  \href{https://doi.org/10.1051/0004-6361/201833910}{\emph{Astron. Astrophys.}
  {\bfseries 641} (2020) A6}
  [\href{https://arxiv.org/abs/1807.06209}{{\ttfamily 1807.06209}}].

\bibitem{DES:2018ekb}
{\scshape DES} collaboration, \emph{{Cosmological Constraints from Multiple
  Probes in the Dark Energy Survey}},
  \href{https://doi.org/10.1103/PhysRevLett.122.171301}{\emph{Phys. Rev. Lett.}
  {\bfseries 122} (2019) 171301}
  [\href{https://arxiv.org/abs/1811.02375}{{\ttfamily 1811.02375}}].

\bibitem{DES:2020mlx}
{\scshape DES} collaboration, \emph{{Dark Energy Survey Year 1 Results:
  Cosmological Constraints from Cluster Abundances, Weak Lensing, and Galaxy
  Correlations}},
  \href{https://doi.org/10.1103/PhysRevLett.126.141301}{\emph{Phys. Rev. Lett.}
  {\bfseries 126} (2021) 141301}
  [\href{https://arxiv.org/abs/2010.01138}{{\ttfamily 2010.01138}}].

\bibitem{Mossa:2020gjc}
V.~Mossa et~al., \emph{{The baryon density of the Universe from an improved
  rate of deuterium burning}},
  \href{https://doi.org/10.1038/s41586-020-2878-4}{\emph{Nature} {\bfseries
  587} (2020) 210}.

\bibitem{Abdalla:2022yfr}
E.~Abdalla et~al., \emph{{Cosmology Intertwined: A Review of the Particle
  Physics, Astrophysics, and Cosmology Associated with the Cosmological
  Tensions and Anomalies}},  in \emph{{2022 Snowmass Summer Study}}, 3, 2022,
  \href{https://arxiv.org/abs/2203.06142}{{\ttfamily 2203.06142}}.

\bibitem{Riess:2019cxk}
A.~G. Riess, S.~Casertano, W.~Yuan, L.~M. Macri and D.~Scolnic, \emph{{Large
  Magellanic Cloud} {Cepheid Standards Provide a 1\% Foundation for the
  Determination of the Hubble Constant} and {Stronger Evidence for Physics}
  beyond {$\Lambda$CDM}},
  \href{https://doi.org/10.3847/1538-4357/ab1422}{\emph{Astrophys. J.}
  {\bfseries 876} (2019) 85}
  [\href{https://arxiv.org/abs/1903.07603}{{\ttfamily 1903.07603}}].

\bibitem{Freedman:2021ahq}
W.~L. Freedman, \emph{{Measurements of the Hubble Constant: Tensions in
  Perspective}},
  \href{https://doi.org/10.3847/1538-4357/ac0e95}{\emph{Astrophys. J.}
  {\bfseries 919} (2021) 16}
  [\href{https://arxiv.org/abs/2106.15656}{{\ttfamily 2106.15656}}].

\bibitem{Riess:2021jrx}
A.~G. Riess et~al., \emph{{A Comprehensive Measurement of the Local Value of
  the Hubble Constant with 1 km s$^{−1}$ Mpc$^{−1}$ Uncertainty from the
  Hubble Space Telescope and the SH0ES Team}},
  \href{https://doi.org/10.3847/2041-8213/ac5c5b}{\emph{Astrophys. J. Lett.}
  {\bfseries 934} (2022) L7}
  [\href{https://arxiv.org/abs/2112.04510}{{\ttfamily 2112.04510}}].

\bibitem{Joudaki:2016mvz}
S.~Joudaki et~al., \emph{{CFHTLenS revisited: assessing concordance with Planck
  including astrophysical systematics}},
  \href{https://doi.org/10.1093/mnras/stw2665}{\emph{Mon. Not. Roy. Astron.
  Soc.} {\bfseries 465} (2017) 2033}
  [\href{https://arxiv.org/abs/1601.05786}{{\ttfamily 1601.05786}}].

\bibitem{Heymans:2020gsg}
C.~Heymans et~al., \emph{{KiDS-1000 Cosmology: Multi-probe weak gravitational
  lensing and spectroscopic galaxy clustering constraints}},
  \href{https://doi.org/10.1051/0004-6361/202039063}{\emph{Astron. Astrophys.}
  {\bfseries 646} (2021) A140}
  [\href{https://arxiv.org/abs/2007.15632}{{\ttfamily 2007.15632}}].

\bibitem{Philcox:2021kcw}
O.~H.~E. Philcox and M.~M. Ivanov, \emph{{BOSS DR12 full-shape cosmology:
  \ensuremath{\Lambda}CDM constraints from the large-scale galaxy power
  spectrum and bispectrum monopole}},
  \href{https://doi.org/10.1103/PhysRevD.105.043517}{\emph{Phys. Rev. D}
  {\bfseries 105} (2022) 043517}
  [\href{https://arxiv.org/abs/2112.04515}{{\ttfamily 2112.04515}}].

\bibitem{Blanchard:2021dwr}
A.~Blanchard and S.~Ili\'c, \emph{{Closing up the cluster tension?}},
  \href{https://doi.org/10.1051/0004-6361/202140974}{\emph{Astron. Astrophys.}
  {\bfseries 656} (2021) A75}
  [\href{https://arxiv.org/abs/2104.00756}{{\ttfamily 2104.00756}}].

\bibitem{DES:2021wwk}
{\scshape DES} collaboration, \emph{{Dark Energy Survey Year 3 results:
  Cosmological constraints from galaxy clustering and weak lensing}},
  \href{https://doi.org/10.1103/PhysRevD.105.023520}{\emph{Phys. Rev. D}
  {\bfseries 105} (2022) 023520}
  [\href{https://arxiv.org/abs/2105.13549}{{\ttfamily 2105.13549}}].

\bibitem{PhysRevD.91.103508}
R.~A. Battye, T.~Charnock and A.~Moss, \emph{Tension between the power spectrum
  of density perturbations measured on large and small scales},
  \href{https://doi.org/10.1103/PhysRevD.91.103508}{\emph{Phys. Rev. D}
  {\bfseries 91} (2015) 103508}.

\bibitem{PhysRevLett.111.161301}
E.~Macaulay, I.~K. Wehus and H.~K. Eriksen, \emph{Lower growth rate from recent
  redshift space distortion measurements than expected from planck},
  \href{https://doi.org/10.1103/PhysRevLett.111.161301}{\emph{Phys. Rev. Lett.}
  {\bfseries 111} (2013) 161301}.

\bibitem{Vagnozzi:2021gjh}
S.~Vagnozzi, \emph{{Consistency tests of \ensuremath{\Lambda}CDM from the early
  integrated Sachs-Wolfe effect: Implications for early-time new physics and
  the Hubble tension}},
  \href{https://doi.org/10.1103/PhysRevD.104.063524}{\emph{Phys. Rev. D}
  {\bfseries 104} (2021) 063524}
  [\href{https://arxiv.org/abs/2105.10425}{{\ttfamily 2105.10425}}].

\bibitem{Schoneberg:2021qvd}
N.~Sch\"oneberg, G.~Franco~Abell\'an, A.~P\'erez~S\'anchez, S.~J. Witte,
  V.~Poulin and J.~Lesgourgues, \emph{{The H0 Olympics: A fair ranking of
  proposed models}},
  \href{https://doi.org/10.1016/j.physrep.2022.07.001}{\emph{Phys. Rept.}
  {\bfseries 984} (2022) 1} [\href{https://arxiv.org/abs/2107.10291}{{\ttfamily
  2107.10291}}].

\bibitem{Gomez-Valent:2021cbe}
A.~G\'omez-Valent, Z.~Zheng, L.~Amendola, V.~Pettorino and C.~Wetterich,
  \emph{{Early dark energy in the pre- and postrecombination epochs}},
  \href{https://doi.org/10.1103/PhysRevD.104.083536}{\emph{Phys. Rev. D}
  {\bfseries 104} (2021) 083536}
  [\href{https://arxiv.org/abs/2107.11065}{{\ttfamily 2107.11065}}].

\bibitem{Anchordoqui:2021gji}
L.~A. Anchordoqui, E.~Di~Valentino, S.~Pan and W.~Yang, \emph{{Dissecting the
  H0 and S8 tensions with Planck + BAO + supernova type Ia in multi-parameter
  cosmologies}},
  \href{https://doi.org/10.1016/j.jheap.2021.08.001}{\emph{JHEAp} {\bfseries
  32} (2021) 28} [\href{https://arxiv.org/abs/2107.13932}{{\ttfamily
  2107.13932}}].

\bibitem{Rashkovetskyi:2021rwg}
M.~Rashkovetskyi, J.~B. Mu\~noz, D.~J. Eisenstein and C.~Dvorkin,
  \emph{{Small-scale clumping at recombination and the Hubble tension}},
  \href{https://doi.org/10.1103/PhysRevD.104.103517}{\emph{Phys. Rev. D}
  {\bfseries 104} (2021) 103517}
  [\href{https://arxiv.org/abs/2108.02747}{{\ttfamily 2108.02747}}].

\bibitem{Bansal:2021dfh}
S.~Bansal, J.~H. Kim, C.~Kolda, M.~Low and Y.~Tsai, \emph{{Mirror twin Higgs
  cosmology: constraints and a possible resolution to the H$_{0}$ and S$_{8}$
  tensions}}, \href{https://doi.org/10.1007/JHEP05(2022)050}{\emph{JHEP}
  {\bfseries 05} (2022) 050}
  [\href{https://arxiv.org/abs/2110.04317}{{\ttfamily 2110.04317}}].

\bibitem{Heisenberg:2022gqk}
L.~Heisenberg, H.~Villarrubia-Rojo and J.~Zosso, \emph{{Can late-time
  extensions solve the H0 and \ensuremath{\sigma}8 tensions?}},
  \href{https://doi.org/10.1103/PhysRevD.106.043503}{\emph{Phys. Rev. D}
  {\bfseries 106} (2022) 043503}
  [\href{https://arxiv.org/abs/2202.01202}{{\ttfamily 2202.01202}}].

\bibitem{Dainotti:2021pqg}
M.~G. Dainotti, B.~De~Simone, T.~Schiavone, G.~Montani, E.~Rinaldi and
  G.~Lambiase, \emph{{On the Hubble constant tension in the SNe Ia Pantheon
  sample}}, \href{https://doi.org/10.3847/1538-4357/abeb73}{\emph{Astrophys.
  J.} {\bfseries 912} (2021) 150}
  [\href{https://arxiv.org/abs/2103.02117}{{\ttfamily 2103.02117}}].

\bibitem{Dainotti:2022bzg}
M.~G. Dainotti, B.~De~Simone, T.~Schiavone, G.~Montani, E.~Rinaldi, G.~Lambiase
  et~al., \emph{{On the Evolution of the Hubble Constant with the SNe Ia
  Pantheon Sample and Baryon Acoustic Oscillations: A Feasibility Study for
  GRB-Cosmology in 2030}},
  \href{https://doi.org/10.3390/galaxies10010024}{\emph{Galaxies} {\bfseries
  10} (2022) 24} [\href{https://arxiv.org/abs/2201.09848}{{\ttfamily
  2201.09848}}].

\bibitem{Murgia:2016ccp}
R.~Murgia, S.~Gariazzo and N.~Fornengo, \emph{{Constraints on the Coupling
  between Dark Energy and Dark Matter from CMB data}},
  \href{https://doi.org/10.1088/1475-7516/2016/04/014}{\emph{JCAP} {\bfseries
  04} (2016) 014} [\href{https://arxiv.org/abs/1602.01765}{{\ttfamily
  1602.01765}}].

\bibitem{Heavens:2017hkr}
A.~Heavens, Y.~Fantaye, E.~Sellentin, H.~Eggers, Z.~Hosenie, S.~Kroon et~al.,
  \emph{{No evidence for extensions to the standard cosmological model}},
  \href{https://doi.org/10.1103/PhysRevLett.119.101301}{\emph{Phys. Rev. Lett.}
  {\bfseries 119} (2017) 101301}
  [\href{https://arxiv.org/abs/1704.03467}{{\ttfamily 1704.03467}}].

\bibitem{Copeland:2006wr}
E.~J. Copeland, M.~Sami and S.~Tsujikawa, \emph{{Dynamics of dark energy}},
  \href{https://doi.org/10.1142/S021827180600942X}{\emph{Int. J. Mod. Phys. D}
  {\bfseries 15} (2006) 1753}
  [\href{https://arxiv.org/abs/hep-th/0603057}{{\ttfamily hep-th/0603057}}].

\bibitem{Clifton:2011jh}
T.~Clifton, P.~G. Ferreira, A.~Padilla and C.~Skordis, \emph{{Modified Gravity
  and Cosmology}},
  \href{https://doi.org/10.1016/j.physrep.2012.01.001}{\emph{Phys. Rept.}
  {\bfseries 513} (2012) 1} [\href{https://arxiv.org/abs/1106.2476}{{\ttfamily
  1106.2476}}].

\bibitem{Li:2011sd}
M.~Li, X.-D. Li, S.~Wang and Y.~Wang, \emph{{Dark Energy}},
  \href{https://doi.org/10.1088/0253-6102/56/3/24}{\emph{Commun. Theor. Phys.}
  {\bfseries 56} (2011) 525} [\href{https://arxiv.org/abs/1103.5870}{{\ttfamily
  1103.5870}}].

\bibitem{Bamba:2012cp}
K.~Bamba, S.~Capozziello, S.~Nojiri and S.~D. Odintsov, \emph{{Dark energy
  cosmology: the equivalent description via different theoretical models and
  cosmography tests}},
  \href{https://doi.org/10.1007/s10509-012-1181-8}{\emph{Astrophys. Space Sci.}
  {\bfseries 342} (2012) 155}
  [\href{https://arxiv.org/abs/1205.3421}{{\ttfamily 1205.3421}}].

\bibitem{PhysRevLett.116.221101}
{\scshape LIGO Scientific and Virgo Collaborations} collaboration, \emph{Tests
  of general relativity with gw150914},
  \href{https://doi.org/10.1103/PhysRevLett.116.221101}{\emph{Phys. Rev. Lett.}
  {\bfseries 116} (2016) 221101}.

\bibitem{Collett:2018gpf}
T.~E. Collett, L.~J. Oldham, R.~J. Smith, M.~W. Auger, K.~B. Westfall, D.~Bacon
  et~al., \emph{{A precise extragalactic test of General Relativity}},
  \href{https://doi.org/10.1126/science.aao2469}{\emph{Science} {\bfseries 360}
  (2018) 1342} [\href{https://arxiv.org/abs/1806.08300}{{\ttfamily
  1806.08300}}].

\bibitem{Planck:2015bue}
{\scshape Planck} collaboration, \emph{{Planck 2015 results. XIV. Dark energy
  and modified gravity}},
  \href{https://doi.org/10.1051/0004-6361/201525814}{\emph{Astron. Astrophys.}
  {\bfseries 594} (2016) A14}
  [\href{https://arxiv.org/abs/1502.01590}{{\ttfamily 1502.01590}}].

\bibitem{tHooft:1993dmi}
G.~'t~Hooft, \emph{{Dimensional reduction in quantum gravity}}, {\emph{Conf.
  Proc. C} {\bfseries 930308} (1993) 284}
  [\href{https://arxiv.org/abs/gr-qc/9310026}{{\ttfamily gr-qc/9310026}}].

\bibitem{Susskind:1994vu}
L.~Susskind, \emph{{The World as a hologram}},
  \href{https://doi.org/10.1063/1.531249}{\emph{J. Math. Phys.} {\bfseries 36}
  (1995) 6377} [\href{https://arxiv.org/abs/hep-th/9409089}{{\ttfamily
  hep-th/9409089}}].

\bibitem{Hawking:1975vcx}
S.~W. Hawking, \emph{{Particle Creation by Black Holes}},
  \href{https://doi.org/10.1007/BF02345020}{\emph{Commun. Math. Phys.}
  {\bfseries 43} (1975) 199}.

\bibitem{Bekenstein:1973ur}
J.~D. Bekenstein, \emph{{Black holes and entropy}},
  \href{https://doi.org/10.1103/PhysRevD.7.2333}{\emph{Phys. Rev. D} {\bfseries
  7} (1973) 2333}.

\bibitem{Bousso:2002ju}
R.~Bousso, \emph{{The Holographic principle}},
  \href{https://doi.org/10.1103/RevModPhys.74.825}{\emph{Rev. Mod. Phys.}
  {\bfseries 74} (2002) 825}
  [\href{https://arxiv.org/abs/hep-th/0203101}{{\ttfamily hep-th/0203101}}].

\bibitem{Bousso:1999xy}
R.~Bousso, \emph{{A Covariant entropy conjecture}},
  \href{https://doi.org/10.1088/1126-6708/1999/07/004}{\emph{JHEP} {\bfseries
  07} (1999) 004} [\href{https://arxiv.org/abs/hep-th/9905177}{{\ttfamily
  hep-th/9905177}}].

\bibitem{Cohen:1998zx}
A.~G. Cohen, D.~B. Kaplan and A.~E. Nelson, \emph{{Effective field theory,
  black holes, and the cosmological constant}},
  \href{https://doi.org/10.1103/PhysRevLett.82.4971}{\emph{Phys. Rev. Lett.}
  {\bfseries 82} (1999) 4971}
  [\href{https://arxiv.org/abs/hep-th/9803132}{{\ttfamily hep-th/9803132}}].

\bibitem{Weinberg:1988cp}
S.~Weinberg, \emph{{The Cosmological Constant Problem}},
  \href{https://doi.org/10.1103/RevModPhys.61.1}{\emph{Rev. Mod. Phys.}
  {\bfseries 61} (1989) 1}.

\bibitem{Sahni:1999gb}
V.~Sahni and A.~A. Starobinsky, \emph{{The Case for a positive cosmological
  Lambda term}}, \href{https://doi.org/10.1142/S0218271800000542}{\emph{Int. J.
  Mod. Phys. D} {\bfseries 9} (2000) 373}
  [\href{https://arxiv.org/abs/astro-ph/9904398}{{\ttfamily
  astro-ph/9904398}}].

\bibitem{Carroll:2000fy}
S.~M. Carroll, \emph{{The Cosmological constant}},
  \href{https://doi.org/10.12942/lrr-2001-1}{\emph{Living Rev. Rel.} {\bfseries
  4} (2001) 1} [\href{https://arxiv.org/abs/astro-ph/0004075}{{\ttfamily
  astro-ph/0004075}}].

\bibitem{Wang:2016och}
S.~Wang, Y.~Wang and M.~Li, \emph{{Holographic Dark Energy}},
  \href{https://doi.org/10.1016/j.physrep.2017.06.003}{\emph{Phys. Rept.}
  {\bfseries 696} (2017) 1} [\href{https://arxiv.org/abs/1612.00345}{{\ttfamily
  1612.00345}}].

\bibitem{Nojiri:2005pu}
S.~Nojiri and S.~D. Odintsov, \emph{{Unifying phantom inflation with late-time
  acceleration: Scalar phantom-non-phantom transition model and generalized
  holographic dark energy}},
  \href{https://doi.org/10.1007/s10714-006-0301-6}{\emph{Gen. Rel. Grav.}
  {\bfseries 38} (2006) 1285}
  [\href{https://arxiv.org/abs/hep-th/0506212}{{\ttfamily hep-th/0506212}}].

\bibitem{Nojiri:2017opc}
S.~Nojiri and S.~D. Odintsov, \emph{{Covariant Generalized Holographic Dark
  Energy and Accelerating Universe}},
  \href{https://doi.org/10.1140/epjc/s10052-017-5097-x}{\emph{Eur. Phys. J. C}
  {\bfseries 77} (2017) 528}
  [\href{https://arxiv.org/abs/1703.06372}{{\ttfamily 1703.06372}}].

\bibitem{Granda:2008dk}
L.~N. Granda and A.~Oliveros, \emph{{Infrared cut-off proposal for the
  Holographic density}},
  \href{https://doi.org/10.1016/j.physletb.2008.10.017}{\emph{Phys. Lett. B}
  {\bfseries 669} (2008) 275}
  [\href{https://arxiv.org/abs/0810.3149}{{\ttfamily 0810.3149}}].

\bibitem{Gao:2007ep}
C.~Gao, F.~Wu, X.~Chen and Y.-G. Shen, \emph{{A Holographic Dark Energy Model
  from Ricci Scalar Curvature}},
  \href{https://doi.org/10.1103/PhysRevD.79.043511}{\emph{Phys. Rev. D}
  {\bfseries 79} (2009) 043511}
  [\href{https://arxiv.org/abs/0712.1394}{{\ttfamily 0712.1394}}].

\bibitem{Nojiri:2021iko}
S.~Nojiri, S.~D. Odintsov and T.~Paul, \emph{{Different Faces of Generalized
  Holographic Dark Energy}},
  \href{https://doi.org/10.3390/sym13060928}{\emph{Symmetry} {\bfseries 13}
  (2021) 928} [\href{https://arxiv.org/abs/2105.08438}{{\ttfamily
  2105.08438}}].

\bibitem{Nojiri:2021jxf}
S.~Nojiri, S.~D. Odintsov and T.~Paul, \emph{{Barrow entropic dark energy: A
  member of generalized holographic dark energy family}},
  \href{https://doi.org/10.1016/j.physletb.2021.136844}{\emph{Phys. Lett. B}
  {\bfseries 825} (2022) 136844}
  [\href{https://arxiv.org/abs/2112.10159}{{\ttfamily 2112.10159}}].

\bibitem{Nojiri:2022dkr}
S.~Nojiri, S.~D. Odintsov and T.~Paul, \emph{{Early and late universe
  holographic cosmology from a new generalized entropy}},
  \href{https://doi.org/10.1016/j.physletb.2022.137189}{\emph{Phys. Lett. B}
  {\bfseries 831} (2022) 137189}
  [\href{https://arxiv.org/abs/2205.08876}{{\ttfamily 2205.08876}}].

\bibitem{Lin_2021}
C.~Lin, \emph{An effective field theory of holographic dark energy},
  \href{https://doi.org/10.1088/1475-7516/2021/07/003}{\emph{Journal of
  Cosmology and Astroparticle Physics} {\bfseries 2021} (2021) 003}.

\bibitem{Heisenberg:2018acv}
L.~Heisenberg, \emph{{Scalar-Vector-Tensor Gravity Theories}},
  \href{https://doi.org/10.1088/1475-7516/2018/10/054}{\emph{JCAP} {\bfseries
  10} (2018) 054} [\href{https://arxiv.org/abs/1801.01523}{{\ttfamily
  1801.01523}}].

\bibitem{Zhang:2009un}
X.~Zhang, \emph{{Holographic Ricci dark energy: Current observational
  constraints, quintom feature, and the reconstruction of scalar-field dark
  energy}}, \href{https://doi.org/10.1103/PhysRevD.79.103509}{\emph{Phys. Rev.
  D} {\bfseries 79} (2009) 103509}
  [\href{https://arxiv.org/abs/0901.2262}{{\ttfamily 0901.2262}}].

\bibitem{Xu:2010gg}
L.~Xu and Y.~Wang, \emph{{Observational Constraints to Ricci Dark Energy Model
  by Using: SN, BAO, OHD, fgas Data Sets}},
  \href{https://doi.org/10.1088/1475-7516/2010/06/002}{\emph{JCAP} {\bfseries
  06} (2010) 002} [\href{https://arxiv.org/abs/1006.0296}{{\ttfamily
  1006.0296}}].

\bibitem{Wang:2011km}
Y.~Wang, L.~Xu and Y.~Gui, \emph{{Probing Ricci dark energy model with
  perturbations by using WMAP seven-year cosmic microwave background
  measurements, BAO and Type Ia supernovae}},
  \href{https://doi.org/10.1103/PhysRevD.84.063513}{\emph{Phys. Rev. D}
  {\bfseries 84} (2011) 063513}
  [\href{https://arxiv.org/abs/1110.4401}{{\ttfamily 1110.4401}}].

\bibitem{Akhlaghi:2018knk}
I.~A. Akhlaghi, M.~Malekjani, S.~Basilakos and H.~Haghi, \emph{{Model selection
  and constraints from Holographic dark energy scenarios}},
  \href{https://doi.org/10.1093/mnras/sty903}{\emph{Mon. Not. Roy. Astron.
  Soc.} {\bfseries 477} (2018) 3659}
  [\href{https://arxiv.org/abs/1804.02989}{{\ttfamily 1804.02989}}].

\bibitem{Hossienkhani:2021emv}
H.~Hossienkhani, N.~Azimi and H.~Yousefi, \emph{{Constraints on the Ricci dark
  energy cosmologies in Bianchi type I model}},
  \href{https://doi.org/10.1142/S021988782150095X}{\emph{Int. J. Geom. Meth.
  Mod. Phys.} {\bfseries 18} (2021) 2150095}.

\bibitem{Najafi:2022wjs}
A.~Najafi and H.~Hossienkhani, \emph{{Using Pantheon and Hubble parameter data
  to constrain the Ricci dark energy in a Bianchi I Universe}},
  \href{https://doi.org/10.1088/1572-9494/ac6ac1}{\emph{Commun. Theor. Phys.}
  {\bfseries 74} (2022) 065401}.

\bibitem{Cid:2020kpp}
A.~Cid, C.~Rodriguez-Benites, M.~Cataldo and G.~Casanova, \emph{{Bayesian
  Comparison of Interacting Modified Holographic Ricci Dark Energy Scenarios}},
  \href{https://doi.org/10.1140/epjc/s10052-021-08841-2}{\emph{Eur. Phys. J. C}
  {\bfseries 81} (2021) 31} [\href{https://arxiv.org/abs/2005.07664}{{\ttfamily
  2005.07664}}].

\bibitem{Malekjani:2018qcz}
M.~Malekjani, M.~Rezaei and I.~A. Akhlaghi, \emph{{Can Holographic dark energy
  models fit the observational data?}},
  \href{https://doi.org/10.1103/PhysRevD.98.063533}{\emph{Phys. Rev. D}
  {\bfseries 98} (2018) 063533}
  [\href{https://arxiv.org/abs/1809.08792}{{\ttfamily 1809.08792}}].

\bibitem{Cardenas:2013moa}
V.~H. Cardenas, A.~Bonilla, V.~Motta and S.~del Campo, \emph{{Constraints on
  Holographic cosmologies from strong lensing systems}},
  \href{https://doi.org/10.1088/1475-7516/2013/11/053}{\emph{JCAP} {\bfseries
  11} (2013) 053} [\href{https://arxiv.org/abs/1310.8251}{{\ttfamily
  1310.8251}}].

\bibitem{PhysRevD.87.043525}
L.~Xu, \emph{Constraints on the holographic dark energy model from type ia
  supernovae, wmap7, baryon acoustic oscillation, and redshift-space
  distortion}, \href{https://doi.org/10.1103/PhysRevD.87.043525}{\emph{Phys.
  Rev. D} {\bfseries 87} (2013) 043525}.

\bibitem{Huang:2012gd}
Z.-P. Huang and Y.-L. Wu, \emph{{Cosmological Constraint and Analysis on
  Holographic Dark Energy Model Characterized by the Conformal-age-like
  Length}}, \href{https://doi.org/10.1142/S0217751X12501308}{\emph{Int. J. Mod.
  Phys. A} {\bfseries 27} (2012) 1250130}
  [\href{https://arxiv.org/abs/1202.3517}{{\ttfamily 1202.3517}}].

\bibitem{Zhang:2012qra}
Z.~Zhang, M.~Li, X.-D. Li, S.~Wang and W.-S. Zhang, \emph{{Generalized
  Holographic Dark Energy and its Observational Constraints}},
  \href{https://doi.org/10.1142/S0217732312501155}{\emph{Mod. Phys. Lett. A}
  {\bfseries 27} (2012) 1250115}
  [\href{https://arxiv.org/abs/1202.5163}{{\ttfamily 1202.5163}}].

\bibitem{Fu:2011ab}
T.-F. Fu, J.-F. Zhang, J.-Q. Chen and X.~Zhang, \emph{{Holographic Ricci dark
  energy: Interacting model and cosmological constraints}},
  \href{https://doi.org/10.1140/epjc/s10052-012-1932-2}{\emph{Eur. Phys. J. C}
  {\bfseries 72} (2012) 1932}
  [\href{https://arxiv.org/abs/1112.2350}{{\ttfamily 1112.2350}}].

\bibitem{PhysRevD.81.083523}
Y.~Wang and L.~Xu, \emph{Current observational constraints to the holographic
  dark energy model with a new infrared cutoff via the markov chain monte carlo
  method}, \href{https://doi.org/10.1103/PhysRevD.81.083523}{\emph{Phys. Rev.
  D} {\bfseries 81} (2010) 083523}.

\bibitem{Li:2009bn}
M.~Li, X.-D. Li, S.~Wang and X.~Zhang, \emph{{Holographic dark energy models: A
  comparison from the latest observational data}},
  \href{https://doi.org/10.1088/1475-7516/2009/06/036}{\emph{JCAP} {\bfseries
  06} (2009) 036} [\href{https://arxiv.org/abs/0904.0928}{{\ttfamily
  0904.0928}}].

\bibitem{Fang:2008sn}
W.~Fang, W.~Hu and A.~Lewis, \emph{{Crossing the Phantom Divide with
  Parameterized Post-Friedmann Dark Energy}},
  \href{https://doi.org/10.1103/PhysRevD.78.087303}{\emph{Phys. Rev. D}
  {\bfseries 78} (2008) 087303}
  [\href{https://arxiv.org/abs/0808.3125}{{\ttfamily 0808.3125}}].

\bibitem{Hogg:2004vw}
D.~W. Hogg, D.~J. Eisenstein, M.~R. Blanton, N.~A. Bahcall, J.~Brinkmann, J.~E.
  Gunn et~al., \emph{{Cosmic homogeneity demonstrated with luminous red
  galaxies}}, \href{https://doi.org/10.1086/429084}{\emph{Astrophys. J.}
  {\bfseries 624} (2005) 54}
  [\href{https://arxiv.org/abs/astro-ph/0411197}{{\ttfamily
  astro-ph/0411197}}].

\bibitem{Marinoni:2012ba}
C.~Marinoni, J.~Bel and A.~Buzzi, \emph{{The Scale of Cosmic Isotropy}},
  \href{https://doi.org/10.1088/1475-7516/2012/10/036}{\emph{JCAP} {\bfseries
  10} (2012) 036} [\href{https://arxiv.org/abs/1205.3309}{{\ttfamily
  1205.3309}}].

\bibitem{Ade:2015hxq}
{\scshape Planck} collaboration, \emph{{Planck 2015 results. XVI. Isotropy and
  statistics of the CMB}},
  \href{https://doi.org/10.1051/0004-6361/201526681}{\emph{Astron. Astrophys.}
  {\bfseries 594} (2016) A16}
  [\href{https://arxiv.org/abs/1506.07135}{{\ttfamily 1506.07135}}].

\bibitem{Ma-Bertschinger:1995asth}
C.-P. Ma and E.~Bertschinger, \emph{Cosmological perturbation theory in the
  synchronous and conformal newtonian gauges},
  \href{https://doi.org//10.1086/176550}{\emph{The Astrophysical Journal}
  {\bfseries 455} (1995) 7}.

\bibitem{Kunz:2006PhyRev}
M.~Kunz and D.~Sapone, \emph{Crossing the phantom divide},
  \href{https://doi.org/https://doi.org/10.1103/PhysRevD.74.123503}{\emph{Physical
  Review D} {\bfseries 74} (2006) 123503}.

\bibitem{BOSS:2016wmc}
{\scshape BOSS} collaboration, \emph{{The clustering of galaxies in the
  completed SDSS-III Baryon Oscillation Spectroscopic Survey: cosmological
  analysis of the DR12 galaxy sample}},
  \href{https://doi.org/10.1093/mnras/stx721}{\emph{Mon. Not. Roy. Astron.
  Soc.} {\bfseries 470} (2017) 2617}
  [\href{https://arxiv.org/abs/1607.03155}{{\ttfamily 1607.03155}}].

\bibitem{2011}
F.~Beutler, C.~Blake, M.~Colless, D.~H. Jones, L.~Staveley-Smith, L.~Campbell
  et~al., \emph{The 6df galaxy survey: baryon acoustic oscillations and the
  local hubble constant},
  \href{https://doi.org/10.1111/j.1365-2966.2011.19250.x}{\emph{Monthly Notices
  of the Royal Astronomical Society} {\bfseries 416} (2011) 3017–3032}.

\bibitem{Ross:2014qpa}
A.~J. Ross, L.~Samushia, C.~Howlett, W.~J. Percival, A.~Burden and M.~Manera,
  \emph{{The clustering of the SDSS DR7 main Galaxy sample \textendash{} I. A 4
  per cent distance measure at $z = 0.15$}},
  \href{https://doi.org/10.1093/mnras/stv154}{\emph{Mon. Not. Roy. Astron.
  Soc.} {\bfseries 449} (2015) 835}
  [\href{https://arxiv.org/abs/1409.3242}{{\ttfamily 1409.3242}}].

\bibitem{Pan-STARRS1:2017jku}
{\scshape Pan-STARRS1} collaboration, \emph{{The Complete Light-curve Sample of
  Spectroscopically Confirmed SNe Ia from Pan-STARRS1 and Cosmological
  Constraints from the Combined Pantheon Sample}},
  \href{https://doi.org/10.3847/1538-4357/aab9bb}{\emph{Astrophys. J.}
  {\bfseries 859} (2018) 101}
  [\href{https://arxiv.org/abs/1710.00845}{{\ttfamily 1710.00845}}].

\bibitem{Arjona:2020yum}
R.~Arjona, J.~Garc\'\i{}a-Bellido and S.~Nesseris, \emph{{Cosmological
  constraints on nonadiabatic dark energy perturbations}},
  \href{https://doi.org/10.1103/PhysRevD.102.103526}{\emph{Phys. Rev. D}
  {\bfseries 102} (2020) 103526}
  [\href{https://arxiv.org/abs/2006.01762}{{\ttfamily 2006.01762}}].

\bibitem{Audren:2012wb}
B.~Audren, J.~Lesgourgues, K.~Benabed and S.~Prunet, \emph{{Conservative
  Constraints on Early Cosmology: an illustration of the Monte Python
  cosmological parameter inference code}},
  \href{https://doi.org/10.1088/1475-7516/2013/02/001}{\emph{JCAP} {\bfseries
  1302} (2013) 001} [\href{https://arxiv.org/abs/1210.7183}{{\ttfamily
  1210.7183}}].

\bibitem{Brinckmann:2018cvx}
T.~Brinckmann and J.~Lesgourgues, \emph{{MontePython 3: boosted MCMC sampler
  and other features}},
  \href{https://doi.org/10.1016/j.dark.2018.100260}{\emph{Phys. Dark Univ.}
  {\bfseries 24} (2019) 100260}
  [\href{https://arxiv.org/abs/1804.07261}{{\ttfamily 1804.07261}}].

\bibitem{Planck:2013pxb}
{\scshape Planck} collaboration, \emph{{Planck 2013 results. XVI. Cosmological
  parameters}},
  \href{https://doi.org/10.1051/0004-6361/201321591}{\emph{Astron. Astrophys.}
  {\bfseries 571} (2014) A16}
  [\href{https://arxiv.org/abs/1303.5076}{{\ttfamily 1303.5076}}].

\bibitem{oliveros2022barrow}
A.~Oliveros, M.~A. Sabogal and M.~A. Acero, \emph{Barrow holographic dark
  energy with {G}randa-{O}liveros cut-off},
  \href{https://doi.org/https://doi.org/10.1140/epjp/s13360-022-02994-z}{\emph{The
  European Physical Journal Plus} {\bfseries 137} (2022) }
  [\href{https://arxiv.org/abs/2203.14464}{{\ttfamily 2203.14464}}].

\bibitem{Smith:2002dz}
{\scshape VIRGO Consortium} collaboration, \emph{{Stable clustering, the halo
  model and nonlinear cosmological power spectra}},
  \href{https://doi.org/10.1046/j.1365-8711.2003.06503.x}{\emph{Mon. Not. Roy.
  Astron. Soc.} {\bfseries 341} (2003) 1311}
  [\href{https://arxiv.org/abs/astro-ph/0207664}{{\ttfamily
  astro-ph/0207664}}].

\bibitem{Abramo:2007iu}
L.~R. Abramo, R.~C. Batista, L.~Liberato and R.~Rosenfeld, \emph{{Structure
  formation in the presence of dark energy perturbations}},
  \href{https://doi.org/10.1088/1475-7516/2007/11/012}{\emph{JCAP} {\bfseries
  11} (2007) 012} [\href{https://arxiv.org/abs/0707.2882}{{\ttfamily
  0707.2882}}].

\bibitem{Abramo:2008ip}
L.~R. Abramo, R.~C. Batista, L.~Liberato and R.~Rosenfeld, \emph{{Physical
  approximations for the nonlinear evolution of perturbations in inhomogeneous
  dark energy scenarios}},
  \href{https://doi.org/10.1103/PhysRevD.79.023516}{\emph{Phys. Rev. D}
  {\bfseries 79} (2009) 023516}
  [\href{https://arxiv.org/abs/0806.3461}{{\ttfamily 0806.3461}}].

\bibitem{PhysRevLett.122.221301}
V.~Poulin, T.~L. Smith, T.~Karwal and M.~Kamionkowski, \emph{Early dark energy
  can resolve the hubble tension},
  \href{https://doi.org/10.1103/PhysRevLett.122.221301}{\emph{Phys. Rev. Lett.}
  {\bfseries 122} (2019) 221301}.

\bibitem{PhysRevD.102.043507}
J.~C. Hill, E.~McDonough, M.~W. Toomey and S.~Alexander, \emph{Early dark
  energy does not restore cosmological concordance},
  \href{https://doi.org/10.1103/PhysRevD.102.043507}{\emph{Phys. Rev. D}
  {\bfseries 102} (2020) 043507}.

\bibitem{Cooke:2017cwo}
R.~J. Cooke, M.~Pettini and C.~C. Steidel, \emph{{One Percent Determination of
  the Primordial Deuterium Abundance}},
  \href{https://doi.org/10.3847/1538-4357/aaab53}{\emph{Astrophys. J.}
  {\bfseries 855} (2018) 102}
  [\href{https://arxiv.org/abs/1710.11129}{{\ttfamily 1710.11129}}].

\bibitem{Colgain:2021beg}
E.~O. Colg\'ain and M.~M. Sheikh-Jabbari, \emph{{A critique of holographic dark
  energy}}, \href{https://doi.org/10.1088/1361-6382/ac1504}{\emph{Class. Quant.
  Grav.} {\bfseries 38} (2021) 177001}
  [\href{https://arxiv.org/abs/2102.09816}{{\ttfamily 2102.09816}}].

\bibitem{Clark:2021hlo}
S.~J. Clark, K.~Vattis, J.~Fan and S.~M. Koushiappas, \emph{{The $H_0$ and
  $S_8$ tensions necessitate early and late time changes to $\Lambda$CDM}},
  \href{https://arxiv.org/abs/2110.09562}{{\ttfamily 2110.09562}}.

\bibitem{Sabla:2022xzj}
V.~I. Sabla and R.~R. Caldwell, \emph{{Microphysics of early dark energy}},
  \href{https://doi.org/10.1103/PhysRevD.106.063526}{\emph{Phys. Rev. D}
  {\bfseries 106} (2022) 063526}
  [\href{https://arxiv.org/abs/2202.08291}{{\ttfamily 2202.08291}}].

\bibitem{Cardona:2022lcz}
W.~Cardona, J.~B. Orjuela-Quintana and C.~A. Valenzuela-Toledo, \emph{{An
  effective fluid description of scalar-vector-tensor theories under the
  sub-horizon and quasi-static approximations}},
  \href{https://doi.org/10.1088/1475-7516/2022/08/059}{\emph{JCAP} {\bfseries
  08} (2022) 059} [\href{https://arxiv.org/abs/2206.02895}{{\ttfamily
  2206.02895}}].

\end{thebibliography}\endgroup

\end{document}